\documentclass[usenatbib]{mn2e}
\usepackage[dvips]{graphicx}
\usepackage{epsfig}
\usepackage{lscape}
\usepackage{amssymb,amsmath}

\newcommand{\kms}{{\rm \,km\,s^{-1}}}
\newcommand{\Mpc}{\,{\rm Mpc}}
\newcommand{\kpc}{\,{\rm kpc}}

\newcommand{\Rcusp}{R_{\rm cusp}}

\title[Line-of-sight lensing and flux anomalies in LCDM]
{On the Effects of Line-of-Sight Structures on Lensing Flux-ratio
Anomalies in a $\Lambda$CDM Universe}

\author[Xu et al.] {D. D. Xu$^{1, 2}$\thanks{E-mail:
    xudd@astro.uni-bonn.de}, Shude Mao$^{1, 3}$,
  Andrew P. Cooper$^{4}$, Liang Gao$^{1}$, Carlos S. Frenk$^{5}$ \and
  Raul E. Angulo$^{4}$, John Helly$^{5}$ \\$^{1}$ National
  Astronomical Observatories, Chinese Academy of Sciences, Beijing,
  100012, China \\$^{2}$ Argelander-Institut f$\ddot{u}$r Astronomie,
  Universit$\ddot{a}$t Bonn, Auf dem H$\ddot{u}$gel 71, 53121 Bonn,
  Germany \\$^{3}$ Jodrell Bank Centre for Astrophysics, the
  University of Manchester, Alan Turing Building, Manchester M13 9PL,
  United Kingdom \\$^{4}$ Max-Planck Institut F\"ur Astrophysik,
  Karl-Schwarzshild-Stra{\ss}e 1, 85740 Garching, Germany \\$^{5}$
  Institute for Computational Cosmology, Dept. of Physics, University
  of Durham, South Road, Durham DH1 3LE, United Kingdom \\
}

\begin{document}

\date{Accepted ...... Received ...... ; in original form......   }

\pagerange{\pageref{firstpage}--\pageref{lastpage}} \pubyear{2011}
\maketitle
\label{firstpage}
\begin{abstract}
  The flux-ratio anomalies observed in multiply-lensed quasar images
  are most plausibly explained as the result of perturbing structures
  superposed on the underlying smooth matter distribution of the
  primary lens. The cold dark matter cosmological model predicts that
  a large number of substructures should survive inside larger halos
  but, surprisingly, this population alone has been shown to be
  insufficient to explain the observed distribution of the flux ratios
  of quasar's multiple images. 
  Other halos (and their subhalos) projected along the line of sight
  to the primary lens have been considered as additional sources of
  perturbation. In this work, %
  we use ray tracing through the Millennium II simulation to
  investigate the importance of projection effects due to halos and
  subhalos of mass $m>10^8 h^{-1}M_{\odot}$ and extend our analysis to
  lower masses, $m\geqslant10^6 h^{-1}M_{\odot}$, using Monte-Carlo
  halo distributions. We find that
  the magnitude of the violation depends strongly on the density
  profile and concentration of the intervening halos, but clustering
  plays only a minor role.
  For a typical lensing geometry (lens at redshift 0.6 and source at
  redshift 2), background haloes (behind the main lens) are more
  likely to cause a violation than foreground halos. We conclude that
  line-of-sight structures can be as important as intrinsic
  substructures in causing flux-ratio anomalies. The combined effect
  of perturbing structures within the lens and along the line of sight
  in the $\Lambda$CDM universe results in a cusp-violation probability
  of 20-30\%.
  This alleviates the discrepancy between models and current data, but
  a larger observational sample is required for a stronger test of the
  theory.

\end{abstract}


\section{INTRODUCTION}

In the cold dark matter (CDM) cosmogony, galaxies are biased tracers
of a filamentary ``cosmic web'' of collapsed regions in the matter
density field -- dark matter haloes. The excellent agreement between
the predictions of this model and observations of the large-scale
clustering of galaxies provides compelling support for CDM. However,
on the scale of individual dark haloes, the model makes a number of
predictions that have yet to be fully verified: cuspy halo density
profiles and a large population of surviving dark matter
substructures. These substructures are the cores of accreted CDM
haloes that persist as long-lived gravitationally bound subhaloes
\citep{Gao2004b,Diemand08Nature,volker08Aq}. Therefore, measurements
of halo density profiles and of the subhalo abundance are crucial
tests of the cosmological model.

Galaxies and their dark matter haloes can act as strong gravitational
lenses, producing distorted and even multiple images of more distant
galaxies and quasars. The distribution and properties of these images
provide sensitive probes of the mass distribution in the lens.  In
some multiply-lensed quasar systems, simple parametric mass models can
fit image positions well, but not their flux ratios. Such anomalies
are interpreted as evidence for complex substructures in lensing
galaxies\footnote{ Apart from quasar images' flux-ratio anomalies,
  another promising method is to use surface brightness anomalies of
  lensed galaxies to identify substructures and constrain their
  properties, see e.g., \citet{VK2009method,VKBTG2009}.}
\citep{MS1998mn,MM2001,MZ2002,Chiba2002,Metcalf2004,Sugai2007,
  McKean2007,More2009,MKA09SubImageAnomaly}.

On scales probed by galactic strong lensing (typically a few
kiloparsecs), predictions from CDM simulations have been compared with
observed flux-ratio anomalies (e.g. \citealt{DK2002,
  BS2004aa,Metcalf2010FluxAnomaly}). Several studies have concluded
that the predicted abundance of dark matter substructures in the
strong-lensing regions of galaxy-sized haloes is not sufficient to
explain the statistics of the currently available sample of known
anomalous lenses \citep{MaoJing04apj, AB06mn, Maccio2006,
  Chen2011CuspViolation}.

In any smooth lens potential producing multiple images of a single
source, a specific magnification ratio (here equivalent to a
flux ratio) of the three most strongly-magnified images will approach
zero asymptotically as the source approaches a cusp of the tangential
caustic. This is known as the ``cusp-caustic relation'' (see
Eq.~\ref{eq:Rcusp} below) (\citealt{BN1986apj, SW1992aa,
  Zakharov1995AA, KGP2003apj}). Structures, either within a lensing
galaxy or projected by chance along the line of sight, will perturb
the potential and alter the flux of one or more images, resulting in a
violation of the cusp-caustic relation. These violations are extreme
cases of flux-ratio anomalies.

\citet{Dandan09AquI,Dandan2010AqII} analyzed flux-ratio distributions
of multiple-imaged background quasars in simulated lensing systems,
using six $\sim10^{12}\mathrm{M_{\sun}}$ CDM haloes and their
substructure populations (subhaloes and streams) from the Aquarius
project (\citealt{volker08Aq}). The effects of baryonic substructures
(satellite galaxies and globular clusters) were also
investigated. These exceptionally high resolution simulations
confirmed that the substructure abundance (in the critical region of a
Milky Way-mass lens) predicted by the CDM model is too low to explain
the observed frequency of cusp-caustic violations.

This apparent deficiency of substructures is not yet a strong
challenge to CDM, in part because the sample of observed lenses is
extremely small.  Furthermore, dark matter haloes and subhaloes are
present along the entire line of sight from the source to the
observer, not just in the lens itself. These independent haloes
projected in front of and behind the lens can also induce
perturbations to the lensing potential and thus cause flux-ratio
anomalies \citep{Chen2003, Wambsganss2005, Metcalf2005a, Metcalf2005b,
  Miranda2007, PunchweinHilbert2009}.

In particular, \citet{Chen2003} used the cross-section (optical depth)
method to calculate the effect of both subhaloes intrinsic to the main
lens and line-of-sight haloes. They found that the former would
dominate the total lensing cross-section, although the exact
percentage was highly sensitive to the spatial distribution of
substructures\footnote{A similar conclusion was reached by
  \citet{Dandan09AquI, Dandan2010AqII}; see also \citet{Anna2011} for
  the observed spatial distribution of luminous satellites in
  early-type galaxies.}; the latter -- line-of-sight haloes, modelled
as singular isothermal spheres -- would contribute to $\leqslant 10\%$
of the total perturbation. \citet{Metcalf2005a} performed ray-tracing
simulations for the line-of-sight lens population ($10^{6}M_{\odot}
\leqslant m \leqslant 10^{9}M_{\odot}$) in a $\Lambda$CDM universe,
which he compared to several observed systems with measured
cusp-caustic ratios. Assuming that haloes have Navarro, Frenk \& White
(NFW) profiles (\citealt{NFW96,NFW97}), Metcalf found that the
predicted abundance of line-of-sight haloes was enough to explain the
observed flux-ratio anomalies of several representative cases.  With a
slightly different approach and assuming haloes to be singular
isothermal spheres, \citet{Miranda2007} found that, with the
additional contribution from haloes along the line of sight, the
observed flux-ratio anomalies can be reproduced with a high
confidence level. \\


In this work, we re-examine the perturbing effect of haloes along the
entire line of sight from the source to the observer by using $N$-body
simulations to generate strong-lensing sight lines and quantify the
flux-ratio distributions for multiply-imaged sources. In $\S$2, we
introduce our method for tracing light deflection through multiple
lens planes. In $\S$3, we present a summary of the cusp-caustic
violations arising from simple perturbation scenarios (varying the
density profiles, angular positions and redshifts of the
perturbers). In $\S$4, we describe our method for generating ``lensing
lighcones'' in the Millennium II $\Lambda$CDM $N$-body simulation
(\citealt{Millennium2}). Results from the analysis of these lensing
cones are given in $\S$5. In $\S$6 we present results using a
Monte-Carlo approach to account for haloes below the mass resolution
limit of Millennium II ($\sim10^{8} h^{-1}M_{\odot}$). Our conclusions
are given in $\S$7. The cosmology of our lensing simulations is the
same as that used for the Millennium-II simulation, with a matter
density $\Omega_{\rm m}= 0.25$, cosmological constant
$\Omega_{\Lambda} = 0.75$, Hubble constant
$h=H_0/(100\kms\Mpc^{-1})=0.73$ and linear fluctuation amplitude
$\sigma_8=0.9$. These values are consistent with cosmological
constraints from the WMAP 1- and 5-year data analyses
(\citealt{WMAP-1, WMAP-5}), but differ at about the 2$\sigma$ level
from more recent WMAP 7-year determinations (\citealt{WMAP-7}).  This
small offset is of no consequence for the topics addressed in this
paper.

\section{Simulations of Light Deflection through Multiple Lens Planes}

\begin{figure}
\includegraphics[width=8cm]{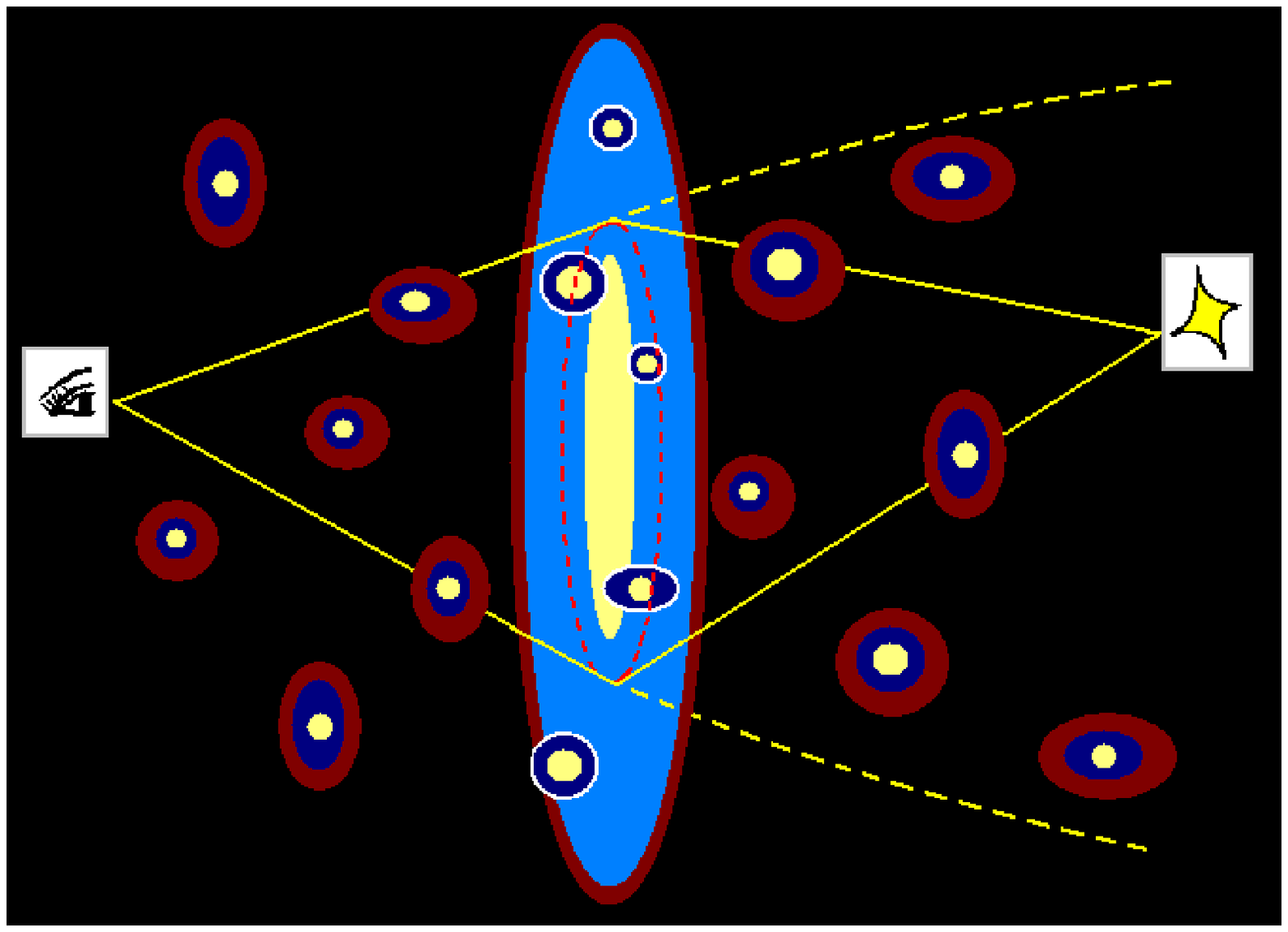}
\includegraphics[width=7.8cm]{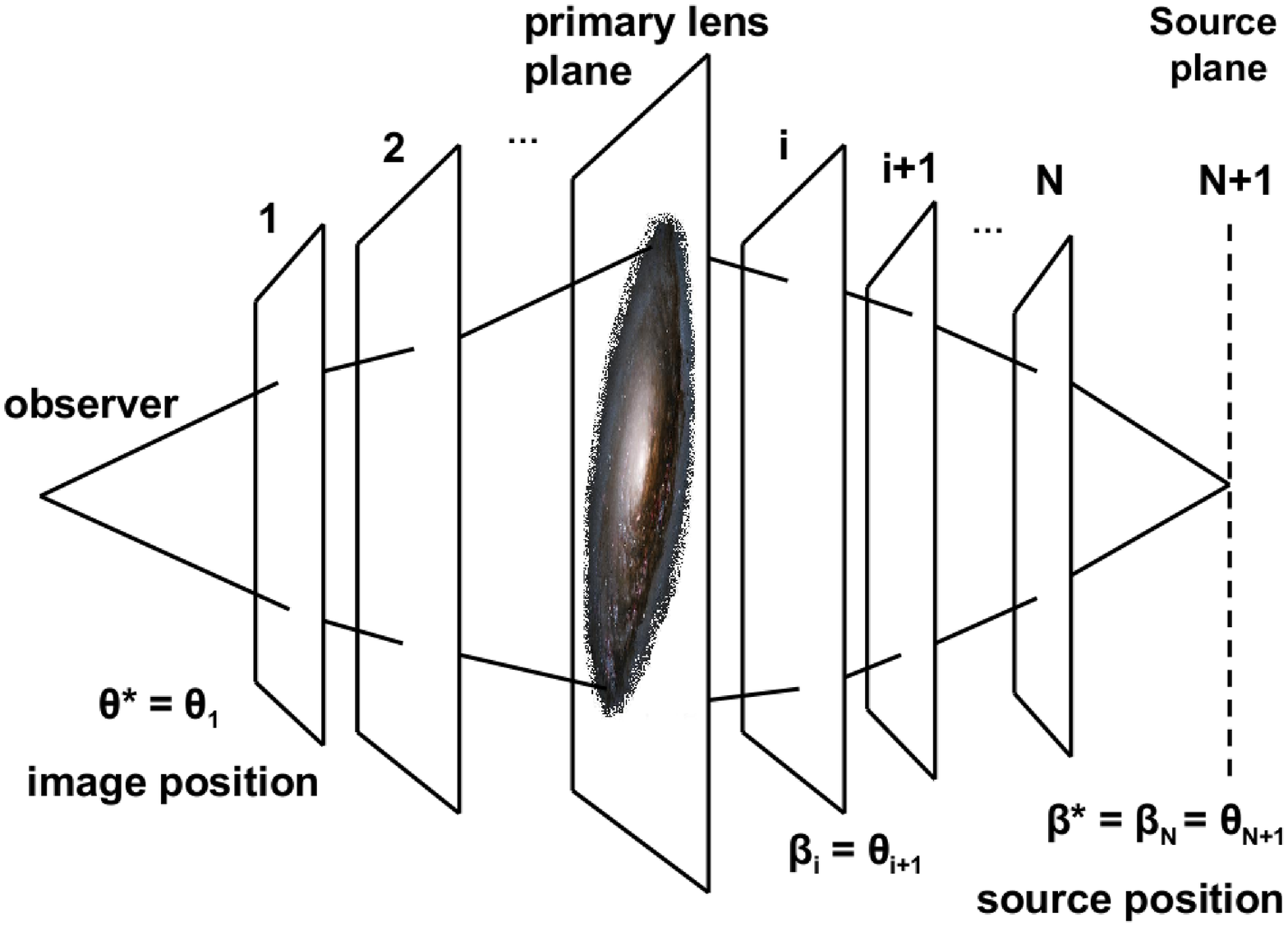}
\caption{An illustration (top) of a light ray propagating through
  intergalactic space from a lensed quasar to an observer. The primary
  galaxy/halo at intermediate redshift causes image splitting due to
  its strong lensing effect. Both intrinsic substructures (satellite
  subhaloes and galaxies) in the primary lens and intergalactic haloes
  along the line of sight perturb the lensing potential and give rise
  to anomalous flux ratios between the images. The corresponding
  illustration of ray tracing through multiple lens planes is given in
  the bottom panel.}
\label{fig:LightDeflection}
\end{figure}

In Chapter 9 of \citet{SchneiderBook1992}, the authors present the
theory of light deflection through multiple lens planes.  As shown in
Fig.~\ref{fig:LightDeflection}, haloes are projected near the line of
sight at all redshifts between the observer and a source at redshift
$z_s$.  The angular position of the source is denoted by
$\vec{\beta}^\star$, and its final image position is denoted by
$\vec{\theta}^\star$. The haloes are assumed to be distributed in $N$
lens planes, each at redshift $z_i$ ($i$=1,2...$N$, and
$z_s=z_{N+1}$). As the light ray passes through each plane, the image
position $\vec{\theta}_{i+1}$ (where the light ray intercepts the
plane) in the ($i+1$)-th lens plane, which is also the source position
$\vec{\beta}_i$ for the $i$-th plane, is related to
$\vec{\beta}^\star$ and $\vec{\theta}^\star$ by the lens equation:
\begin{equation}
\vec{\theta}_{i+1}=\vec{\beta}_i=\vec{\theta}_1-\sum^i_{j=1}
\frac{D_{j,i+1}}{D_{i+1}} \vec{\hat{\alpha}}_j(\vec{\theta}_j),
\label{eq:LensEq}
\end{equation}
where $\vec{\theta}_{N+1}=\vec{\beta}_N=\vec{\beta}^\star$, and
$\vec{\theta}_1=\vec{\theta}^\star$.
$\vec{\hat{\alpha}}_j(\vec{\theta}_j)$ is the deflection angle a
light ray undergoes in the $j$-th plane at $\vec{\theta}_j$.
$D_{i+1}$ and $D_{j,i+1}$ are angular diameter distances between the
($i+1$)-th plane and the observer, and between the ($i+1$)-th plane
and the $j$-th plane, respectively. $D_{N+1}=D_s$ is the angular
diameter distance of the source. The Jacobian matrix $A_i$ of the
mapping between $\vec{\theta}_1$ and the source position
$\vec{\beta}_i$ for the $i$-th plane is given by:
\begin{equation}
A_{i+1}=\frac{\partial{\vec{\beta}_i}}{\partial{\vec{\theta}_1}}
=\frac{\partial{\vec{\theta}_{i+1}}}{\partial{\vec{\theta}_1}}
= I - \sum^i_{j=1} \frac{D_{j,i+1}}{D_{i+1}}
\frac{\partial{\vec{\hat{\alpha}}_j}}{\partial{\vec{\theta}_j}}
\frac{\partial{\vec{\theta}_j}}{\partial{\vec{\theta}_1}}
\label{eq:Jacobian}
\end{equation}
and $A_N\equiv A_{\rm s} =
\frac{\partial{\vec{\beta}_N}}{\partial{\vec{\theta}_1}}$ is the
overall Jacobian matrix, describing the mapping relation between
$\vec{\beta}^\star$ and $\vec{\theta}^\star$. 

Images of any given background source can be accurately and
efficiently identified using the Newton-Raphson method, once the
mapping relation (the overall Jacobian matrix $A_{\rm
  s}=\frac{\partial{\vec{\beta}_N}}{\partial{\vec{\theta}_1}}$) is
obtained.  We determine the Jacobian matrix numerically, as
follows. For each lens plane, a rigid grid of $1000\times1000$ is
applied to cover a central region of $5\arcsec\times 5\arcsec$. Source
positions $\vec{\beta}_N(\vec{\theta}_1)$ that correspond to grid
points $\vec{\theta}_1$ in the first lens plane (which is also the
final image plane) are calculated through the multi-plane lens
equation (Eq.~\ref{eq:LensEq}). An arbitrary light ray propagating
through a lens plane will not necessarily hit a grid point of the mesh
that covers that plane. Therefore, for any given position
$\vec{\theta}_i$ of the $i$-th lens plane, the deflection angle
$\vec{\hat{\alpha}}_i(\vec{\theta}_i)$ is obtained by linear
interpolation using the values for the four nearest grid points. Once
$\vec{\beta}_N(\vec{\theta}_1)$ is obtained, the Jacobian matrix
$A_{\rm s}(\vec{\theta}_1)
(={\partial{\vec{\beta}_N}}/{\partial{\vec{\theta}_1}})$ for the image
plane grid points $\vec{\theta}_1$ can be derived using finite
differencing with the five-point stencil method (accurate to the third
order). Again, linear interpolation is used to obtain each element of
the Jacobian matrix, $A_{\rm s}(\vec{\theta})$, at any given image
position $\vec{\theta}$. The corresponding image magnification
$\mu(\vec{\theta})$ is then given by $\mu = {\rm det}A_{\rm s}^{-1}$.

We use our multi-plane ray tracing code with a resolution of
0.005$\arcsec$ per pixel in the lens and image planes, which we find
sufficient to accurately reproduce the lensing properties of a number
of simple analytical cases.

\section{Anomalous Flux Ratios and Cusp-Caustic Violations}

\begin{figure}
\includegraphics[width=8cm]{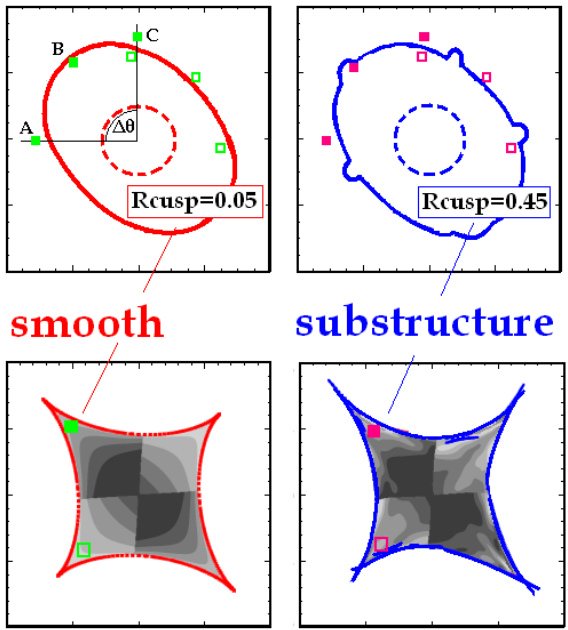}
\caption{An illustration of how the presence of substructures affects
  the cusp-caustic relation. The upper panels show the critical curves
  in the image plane; the bottom panels are the contour maps of
  $\Rcusp$ for sources within the tangential caustic in the source
  plane. Squares indicate positions of close triple images and the
  corresponding sources in the two planes. The image opening angle
  $\Delta\theta$ is labelled for one case in the top left panel. The
  left column shows cases with a smooth lens potential. In the right
  column, we show cases where substructures are present. The
  cusp-caustic relation is violated when a perturbing structure is
  projected near the image positions around the critical curve (see
  text).}
\label{fig:Sub_Rcusp_ilus}
\end{figure}

The cusp-caustic relation (\citealt{BN1986apj}; \citealt{SW1992aa};
\citealt{Zakharov1995AA}; \citealt{KGP2003apj}) is defined as:
\begin{equation}
  \Rcusp \equiv \frac{|\mu_A + \mu_B +
    \mu_C|}{|\mu_A|+|\mu_B|+|\mu_C|} \rightarrow 0
\label{eq:Rcusp}
\end{equation}
when $\mu_{\rm total}=|\mu_{\rm A}|+|\mu_{\rm B}|+|\mu_{\rm C}|
\rightarrow \infty$. $\mu$ denote the magnifications of the three
closest images ($A$, $B$ and $C$) of a background point source located
near a cusp of the tangential caustic (as shown in
Fig.~\ref{fig:Sub_Rcusp_ilus}). Observationally, source positions
cannot be directly measured -- instead, an image opening angle is
often used as an indicator of the proximity of a source to the nearest
cusp of the tangential caustic. This opening angle $\Delta\theta$ is
measured between lines joining the centre of the lens to the two outer
images $A$ and $C$. As the source moves outwards (towards the nearest
cusp), $\Delta\theta \rightarrow 0$, $\mu_{\rm total} \rightarrow
\infty$, and $\Rcusp$ will go to zero asymptotically. This
relationship holds for any smooth lens potential.

Fig.~\ref{fig:Sub_Rcusp_ilus} illustrates how perturbing structures
change the cusp-caustic relation. The upper panels show the critical curves in
the image plane, and the bottom panels are contour maps of $\Rcusp$ for
sources within the tangential caustic in the source plane. Left and right
columns show smooth lens potentials and lens potentials with substructures,
respectively. Substructures located near the critical curve will affect
images nearby and result in significantly larger values of
$\Rcusp$, violating the predicted ratios of image magnifications (fluxes)
given by Eq.~(\ref{eq:Rcusp}).

\subsection{Observational samples}

Multiple images of lensed quasars with small $\Delta\theta$ are ideal
cases to examine violations of the cusp-caustic relation and can be
used to put constraints on the properties of perturbing structures.
This is especially true when their fluxes are measured in the radio
and mid-infrared, as the interpretation of optical and near-infrared
flux ratios is complicated by stellar microlensing and dust
extinction. At the present time, only five cusp-geometry lensing
systems with image opening angle $\Delta\theta \leqslant 90^{\circ}$
are known. These were used for statistical comparisons to the
simulations in our previous work (\citealt{Dandan09AquI,
  Dandan2010AqII}). All five cases have surprisingly large $\Rcusp$
values which are difficult to explain with simple smooth lens
models. Of these five (flux-ratio measurements), two that were
obtained in the optical have been proven to be affected by
microlensing; the other three were from the CLASS survey
(\citealt{Browne2003,Myers2003}) at radio wavelengths and are thought
to be more secure cases of perturbations due to substructures in the
lens.

\begin{table}
\centering \caption{Four-image quasar lensing systems with
$\Delta\theta \leqslant 120^{\circ}$ measured at radio wavelengths
(CLASS).} \small\addtolength{\tabcolsep}{-1.5pt} \label{tab:obs120}
\begin{minipage} {\textwidth}
\begin{tabular}[b]{l|c|c|l}\hline
Systems & $\Delta\theta~({\circ})$ & $\Rcusp$ & Reference \\ \hline
(1) B2045+265 & 34.9 & 0.501 & \citet{Fassnacht1999B2045} \\
& & & \citet{Koopmans2003JVASCLASS}\\
(2) B0712+472 & 76.9 & 0.255 & \citet{Jackson98B0712} \\
& & & \citet{Koopmans2003JVASCLASS} \\
(3) B1422+231 & 77.0 & 0.187 & \citet{Patnaik1999B1422} \\
& & & \citet{Koopmans2003JVASCLASS}\\
(4) MG0414+053 & 101.5 & 0.227 & \citet{Hewitt1992MG0414} \\ 
& & & \citet{Katz1997MG0414}\\
(5) B1555+375 & 102.6 & 0.417 & \citet{Marlow1999B1555} \\
& & & \citet{Koopmans2003JVASCLASS}\\
\hline
\end{tabular}
\end{minipage}
\end{table}

Table 2 of \citet{Chen2011CuspViolation} lists all of the currently
observed $\Rcusp$-$\Delta\theta$ pairs for systems with four distinct
point-like images of quasars lensed by one single galaxy. Our
Table~\ref{tab:obs120} lists those with their flux ratios measured in
the radio and image opening angles $\Delta\theta \leqslant
120^{\circ}$. We compare our simulations with this observational
sample of lenses, with no additional selection criteria.

Notice that four of the five lenses listed in Table 1, namely,
B0712+472, B1422+231, B2045+265 and MG0414+053, are reported as having
visible companions (satellites/group galaxies) projected near the main
lensing galaxies (\citealt{Fassnacht02B0712Group,
  HB94B1422,Grant04B1422Group,McKean2007,Falco97MG0414}). Fitting the
observed image positions using a singular isothermal ellipsoidal model
yields velocity dispersions $\sigma$ ranging from $200-400$ km/s, and
axis ratios $q$ between $0.75-0.9$ (\citealt{Sluse2012COSMOGRAIL25}),
except for B1555, which requires $\sigma=133$ km/s and $q=0.45$
(\citealt{Marlow1999B1555}).

\subsection{Statistical measures for the cusp-caustic violation:
  $P(\geqslant \Rcusp | \Delta\theta \pm2.5^{\circ})$ and
  $P^{90}(\Rcusp^{0.187})$}

Given a simulated lensing system, we compare to the observations in
Table 1 by generating a large number of realisations of background
sources with $\Delta\theta \leqslant 120^{\circ}$. We calculate
$\Rcusp$ for each realisation and evaluate $P(\geqslant \Rcusp |
\Delta\theta \pm2.5^{\circ})$ for this ensemble of realisations. This
is defined as the probability for $\Rcusp$, measured for sources with
image opening angles $\in
[\Delta\theta-2.5^{\circ},~\Delta\theta+2.5^{\circ}]$ (i.e.  within a
five-degree opening-angle span centred at $\Delta\theta$) to be larger
than a particular threshold value. Lenses with more perturbations will
result in large $\Rcusp$ values for many source positions and thus
have a higher $P(\geqslant \Rcusp | \Delta\theta \pm2.5^{\circ})$ than
lenses with fewer perturbations.

\begin{figure}
\begin{center}
\includegraphics[width=6.5cm]{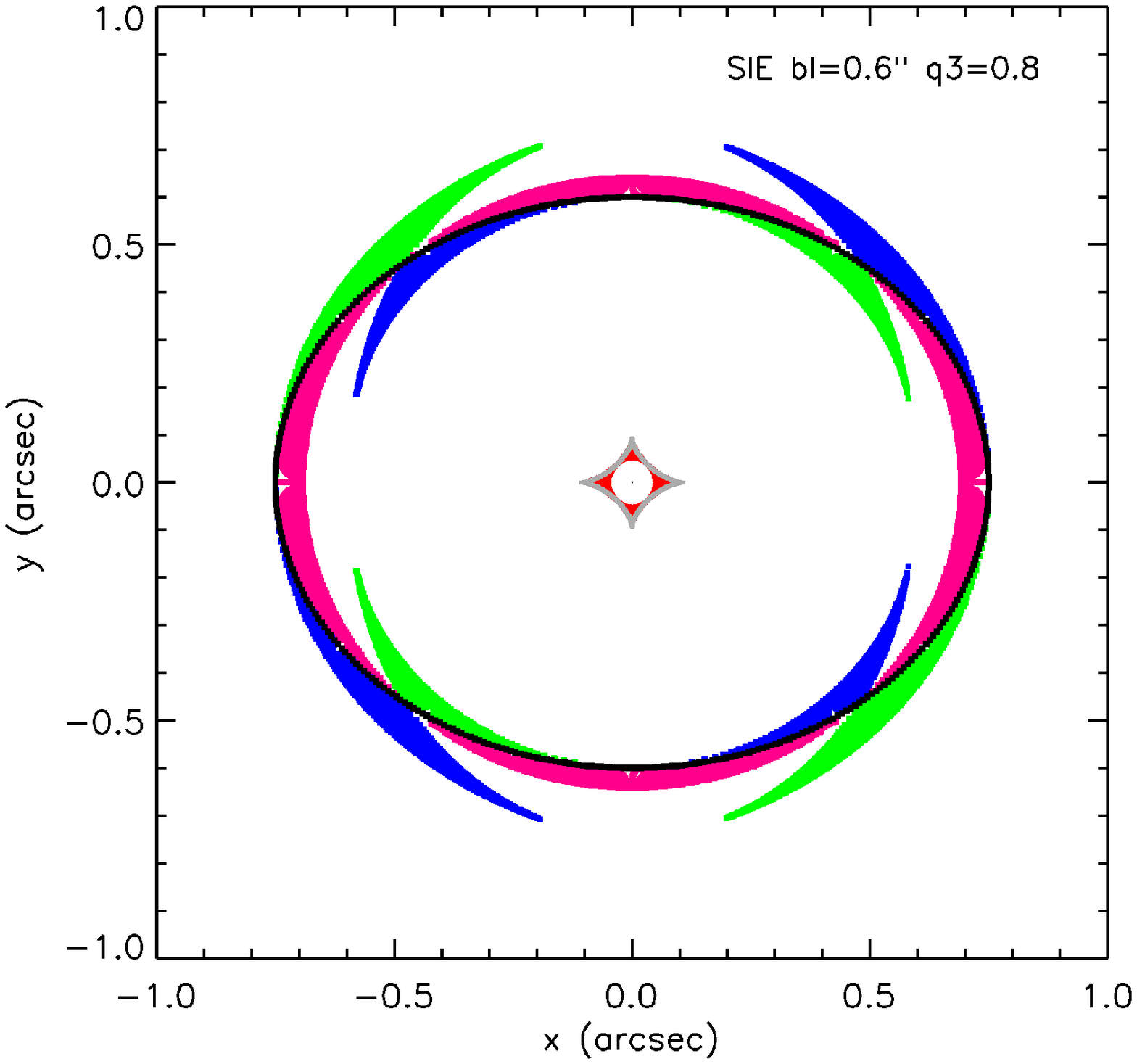} \\
\includegraphics[width=6.5cm]{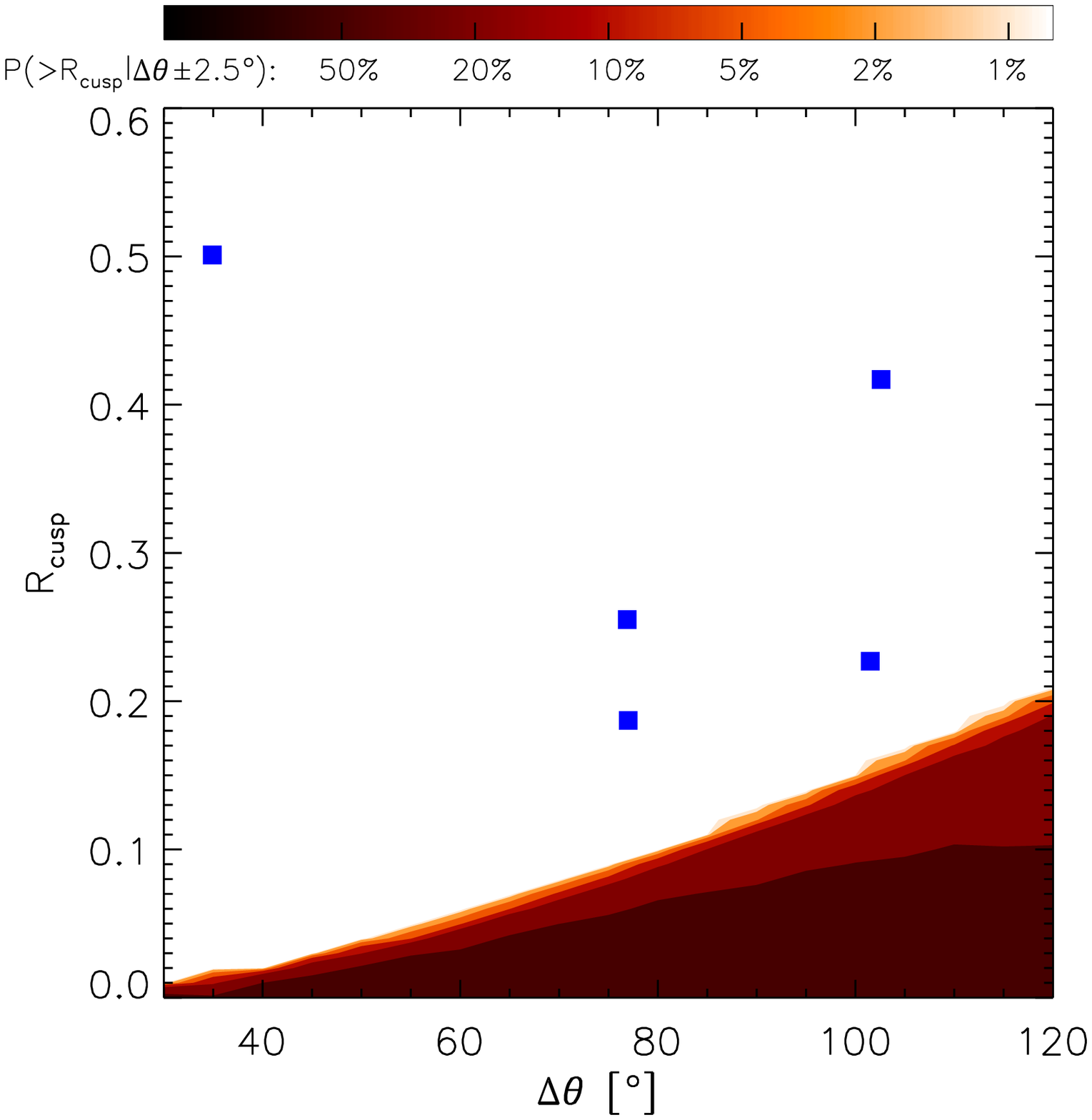} \\
\includegraphics[width=6.5cm]{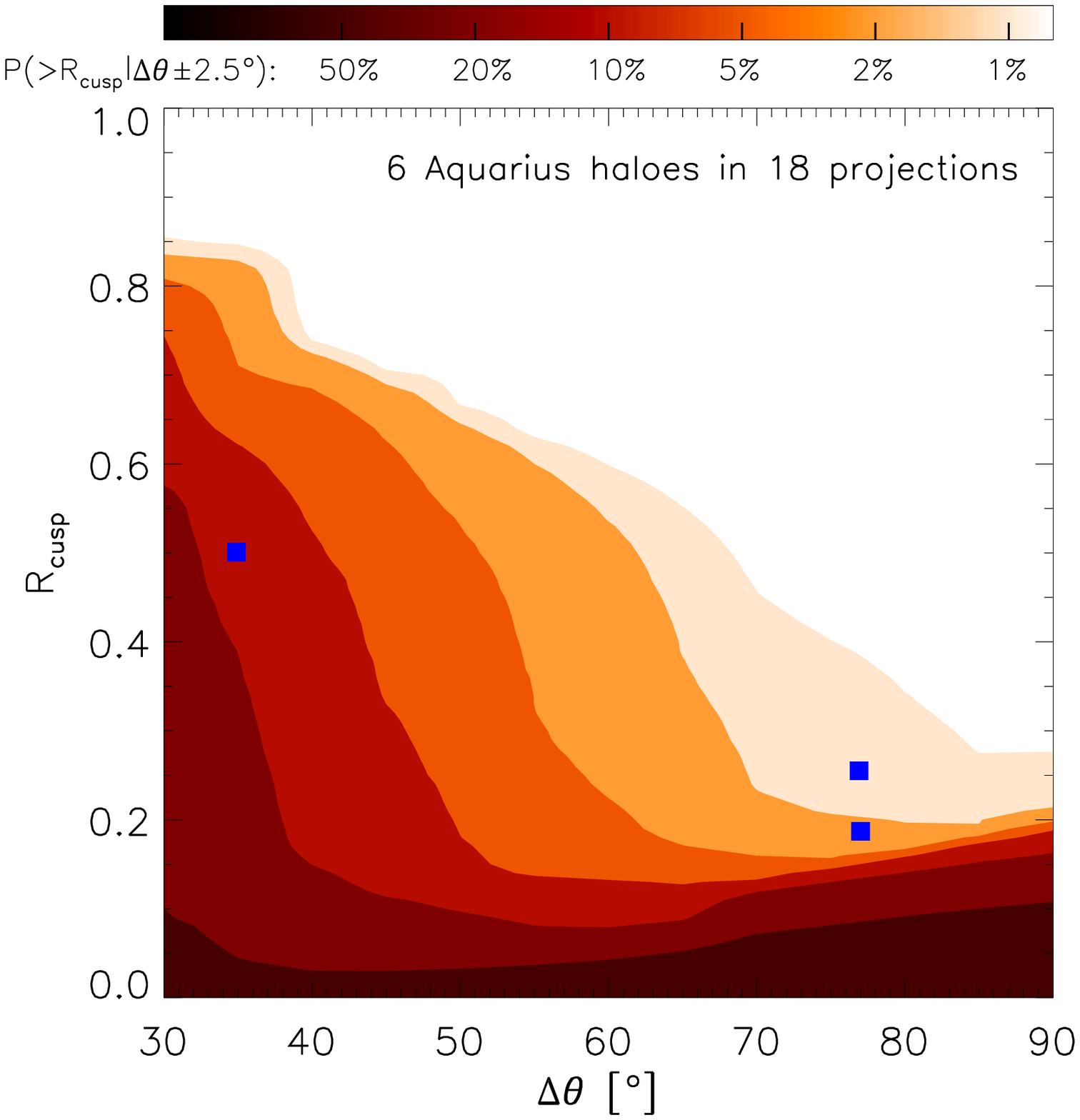}
\end{center}

\caption{Top panel: Close triple image configurations for a SIE lens
  with $b_I=0.6\arcsec$ and $q_3=0.8$. The critical curve of the lens
  is shown in black and its caustics in grey. The regions sampled by
  ``cusp sources'' are shown in red, and the corresponding
  distributions of the three images are shown as green, pink and blue
  regions around the critical curve. Middle panel: the corresponding
  probability contour map for $P(\geqslant \Rcusp | \Delta\theta
  \pm2.5^{\circ})$. Contour levels of 1\%, 2\%, 5\%, 10\%, 20\% and
  50\% (from top to bottom) are plotted. Blue squares are the five
  radio measurements (with $\Delta\theta \leqslant 120^{\circ}$) so
  far available. Bottom panel: the average probability contour map
  (for $\Delta\theta \leqslant 90^{\circ}$) of violations due to
  substructures in the Aquarius haloes (using results from
  \citealt{Dandan09AquI}).} \label{fig:SIEandAq}
\end{figure}

We illustrate our use of the $P(\geqslant \Rcusp | \Delta\theta
\pm2.5^{\circ})$ in Fig.~\ref{fig:SIEandAq}. The top panel shows a
typical example of a close triple image configuration for cusp sources
with $\Delta\theta \leqslant 120^{\circ}$. In this case, the lensing
galaxy has a (smooth) singular isothermal ellipsoidal (SIE) profile
(see \citet{KKprofile1998} for notations of $b_I$, $b_{\rm SIE}$,
$q_3$, and $s_0$ herebelow) with lensing strength $b_I=0.6\arcsec$ and
axis ratio $q_3=0.8$, and is located at redshift $z_d=0.6$; the source
redshift is $z_s=2$. The corresponding contour map of $P(\geqslant
\Rcusp | \Delta\theta \pm2.5^{\circ})$ in the $\Rcusp$-$\Delta\theta$
plane is given in the middle panel. Also plotted are the radio
measurements for the currently best available sample (listed in
Table~\ref{tab:obs120}). These are clearly
inconsistent with the smooth-lens $\Rcusp$ distribution. \\

When we include the substructures within the lensing galaxy and its
dark matter halo, the regular $\Rcusp$ distribution for a smooth lens
potential disappears. The bottom panel in Fig.~\ref{fig:SIEandAq}
shows the average distribution of $P(\geqslant \Rcusp | \Delta\theta
\pm2.5^{\circ})$ when including the subhalo population from the
Aquarius simulations (\citealt{Dandan09AquI}). At small
$\Delta\theta$, violations are more significant than on larger
scales. The smallest $\Rcusp$ measured among all observed cusp-caustic
systems is 0.187 (from B1422). In \citet{Dandan09AquI}, we calculated
$P^{90}(\Rcusp^{0.187})$, which is the probability for $\Rcusp$ to be
larger than or equal to 0.187, computed over all realizations with
$\Delta\theta \leqslant 90^{\circ}$. $P^{90}(\Rcusp^{0.187})$ was
found to be $\sim$10\%. We concluded that it is difficult to explain
the observed $\Rcusp$ distribution (especially at larger
$\Delta\theta$) with a subhalo population similar to that produced in
the Aquarius simulations. This motivates the search for other sources
of perturbations to the lens potential.

In this work, we use $P^{90}(\Rcusp^{0.187})$ as an overall estimate
for the probability of observing cusp-caustic violations, in order to
compare with our previous work. The value of $90^{\circ}$ is chosen in
order to compare with \citet{AB06mn}, who adopted $\Delta\theta
\leqslant 90^{\circ}$ to select cusp-like lenses which can be best
used to test the cusp-caustic violation. Varying the upper limit of
$\Delta\theta$ will change the probability of cases with
$\Rcusp\geqslant0.187$. However, our final conclusion (in $\S$7) is
based on a statistical argument that only applies to the observed
lenses with $\Delta\theta \leqslant 90^{\circ}$.

\subsection{Simple perturbation scenarios with different halo redshifts, masses, profiles and concentrations}

A number of parameters determine the importance of these perturbers
for creating violations to the cusp-caustic relation: most significant
are their masses, density profiles, redshifts and impact parameters to
the line of sight. Before presenting the results from general lines of
sight taken from $N$-body simulations, we first show several simple
perturbation scenarios to illustrate, individually, the effects of
these different parameters, in the case of a single perturbing halo.

\begin{figure*}
\begin{center}
\includegraphics[width=8.5cm]{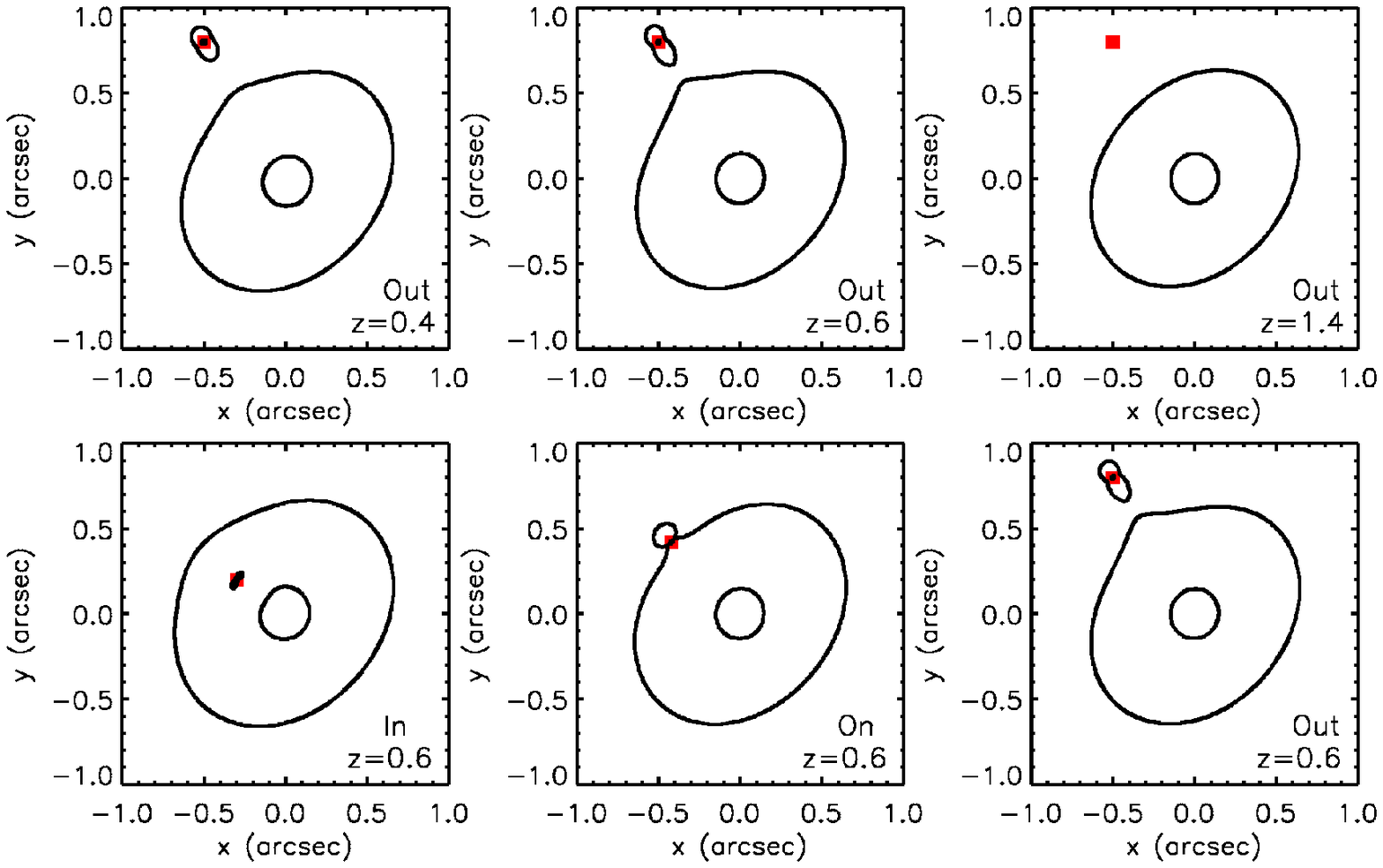}
\includegraphics[width=8.5cm]{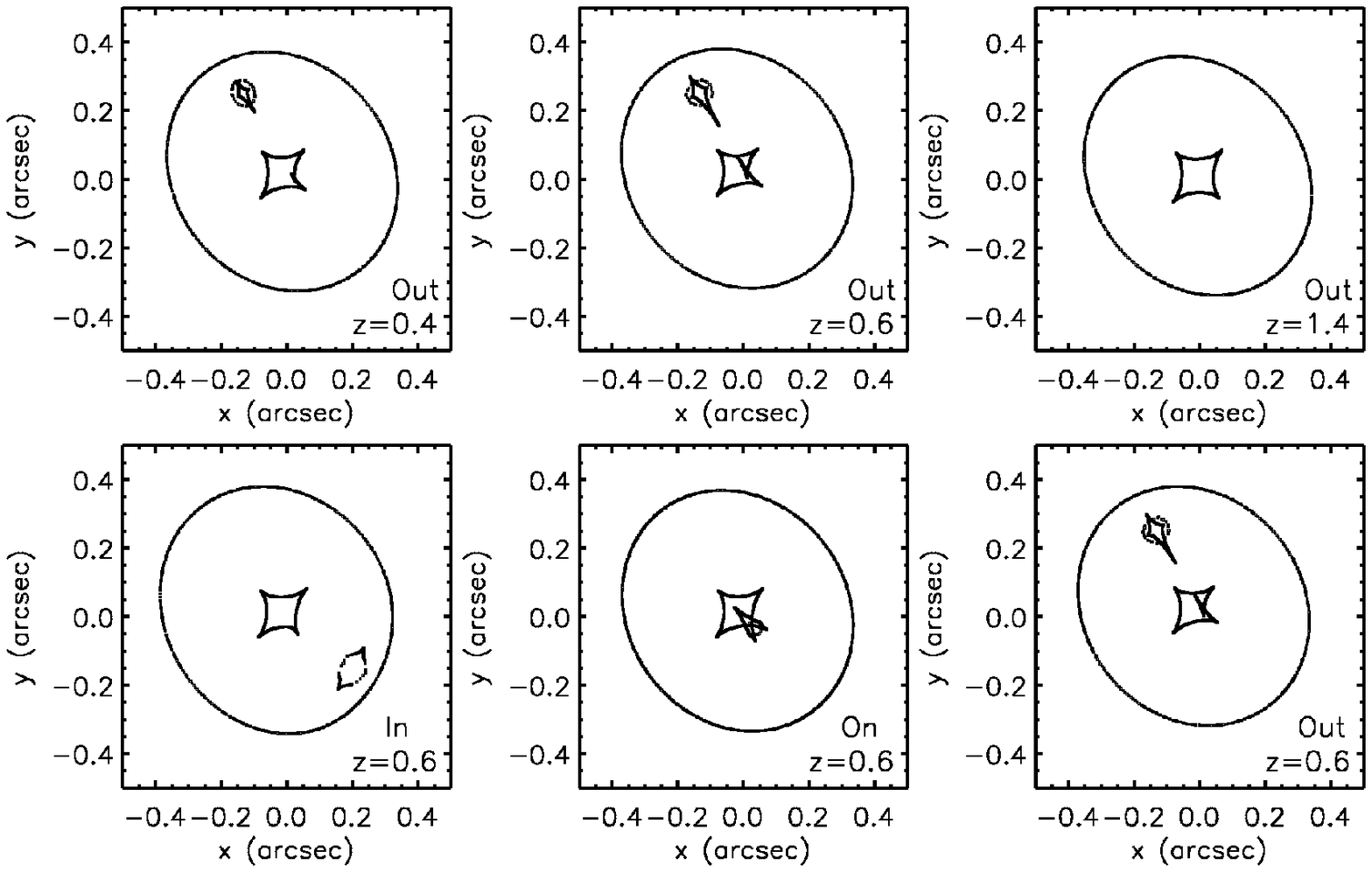}
\end{center}
\caption{Critical curves (six panels on the left) and caustics (six
  panels on the right) produced by the same smooth lens potential plus
  a $m=10^{10} M_{\odot}$ perturber (indicated by red squares),
  modelled by a truncated singular isothermal sphere. Top row: the
  perturber's angular position is fixed and its redshift set at
  $z=0.4$ (in the foreground, left panel), $z=0.6$ (in the primary
  lens plane, middle panel) and $z=1.4$ (in the background, right
  panel). Bottom row: The redshift of the perturber is fixed ($z=0.6$)
  and its impact parameter is changed from inside the tangential
  critical curve (left panel, labelled as ``In''), to overlapping
  (middle panel, labelled as ``On'') to outside (right panel, labelled
  as ``Out''). Parameters are noted in each panel. Note the wiggles
  induced in the critical curves and the production of secondary
  critical curves and caustics.}.
\label{fig:CCZpos} \end{figure*}

Fig.~\ref{fig:CCZpos} shows critical curves and caustics produced by a
main-lens potential of an isothermal ellipsoid with $b_I=0.6\arcsec$,
$q_3=0.8$ and core size $s_0=0.05\arcsec$, located at $z_d=0.6$ for a
source at redshift $z_s=2$, plus a perturber of $m=10^{10} M_{\odot}$
modelled with a truncated singular isothermal sphere. The panels in
this figure correspond to different scenarios. In the upper row the
perturber's angular position is fixed (outside the tangential critical
curve), and we vary its redshift: $z=0.4$ (foreground), $z=0.6$ (in
the main-lens plane) and $z=1.4$ (background). In the lower row, we
fix the redshift of the perturber to the main lens plane ($z=0.6$) and
change its impact parameter, such that it is projected within, on top
of and outside the tangential critical curve (left to right panels,
respectively). Wiggles and swallow tails are introduced to the
critical lines and caustics by the added perturbing structure; massive
perturbers (or those with compact density profiles) can even cause a
secondary set of criticals and caustics. Images located around these
wiggle features violate the cusp-caustic relation most strongly.

\begin{figure*}
\includegraphics[width=5.5cm]{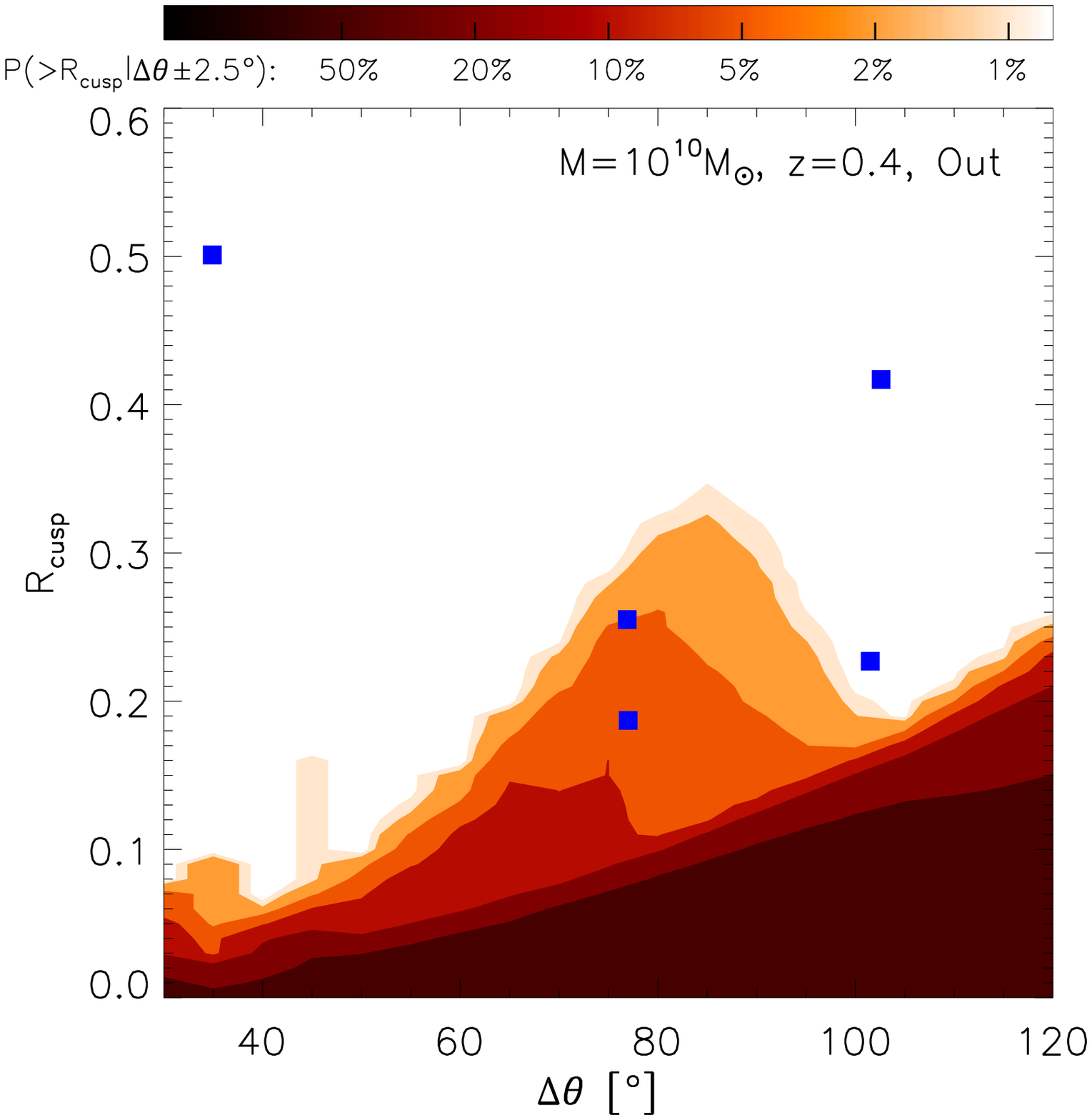}
\includegraphics[width=5.5cm]{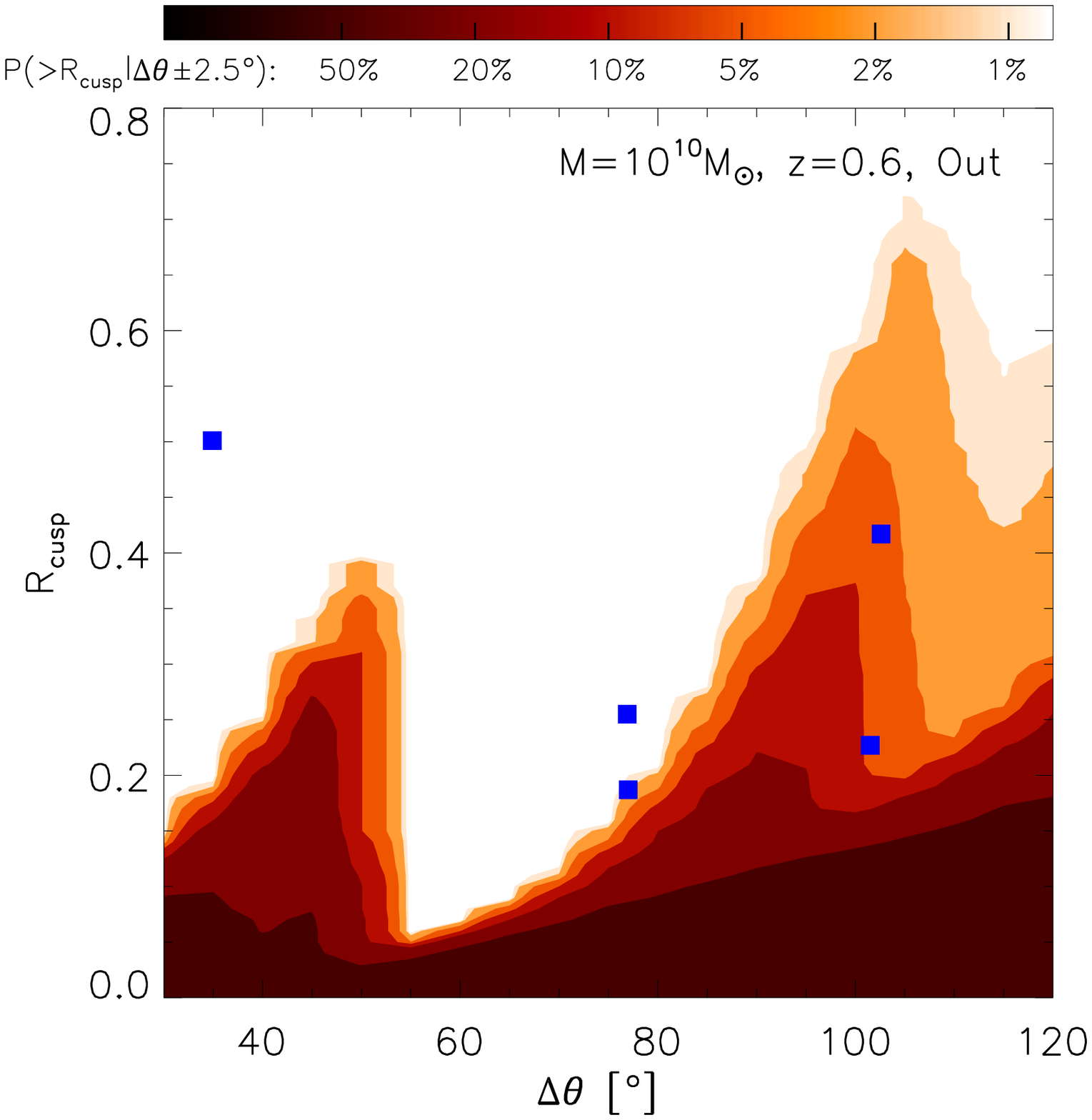}
\includegraphics[width=5.5cm]{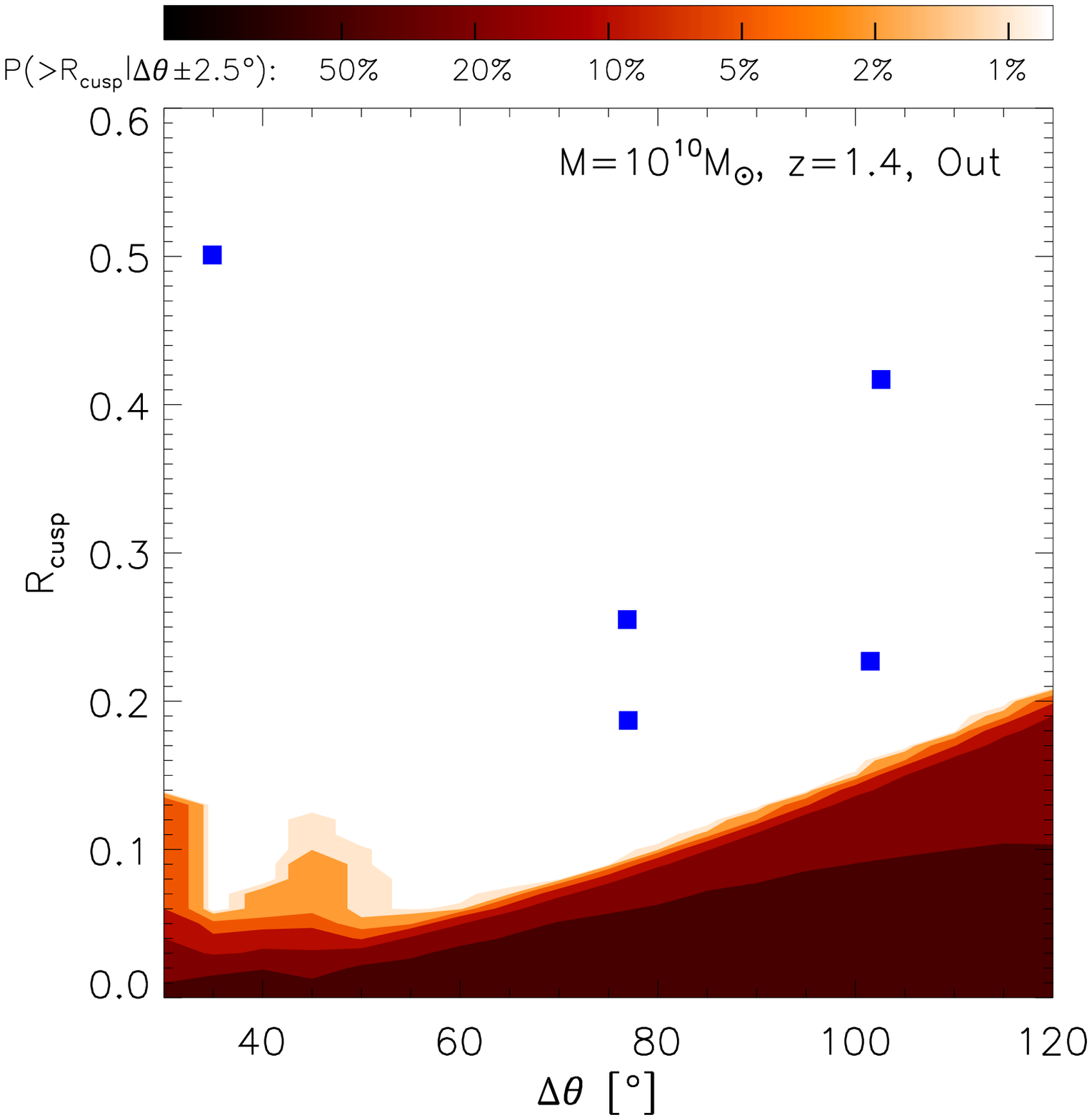} \\
\includegraphics[width=5.5cm]{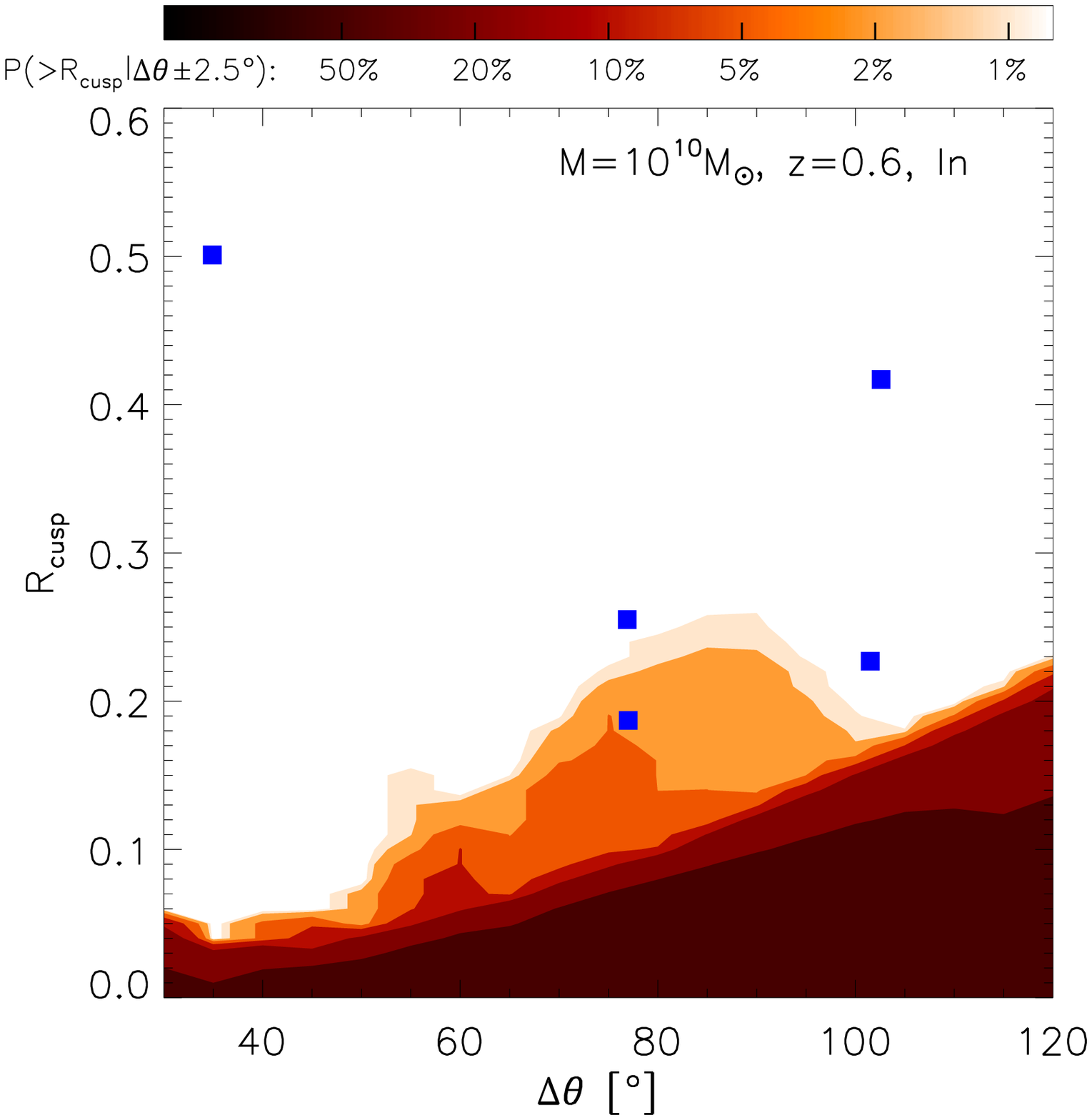}
\includegraphics[width=5.5cm]{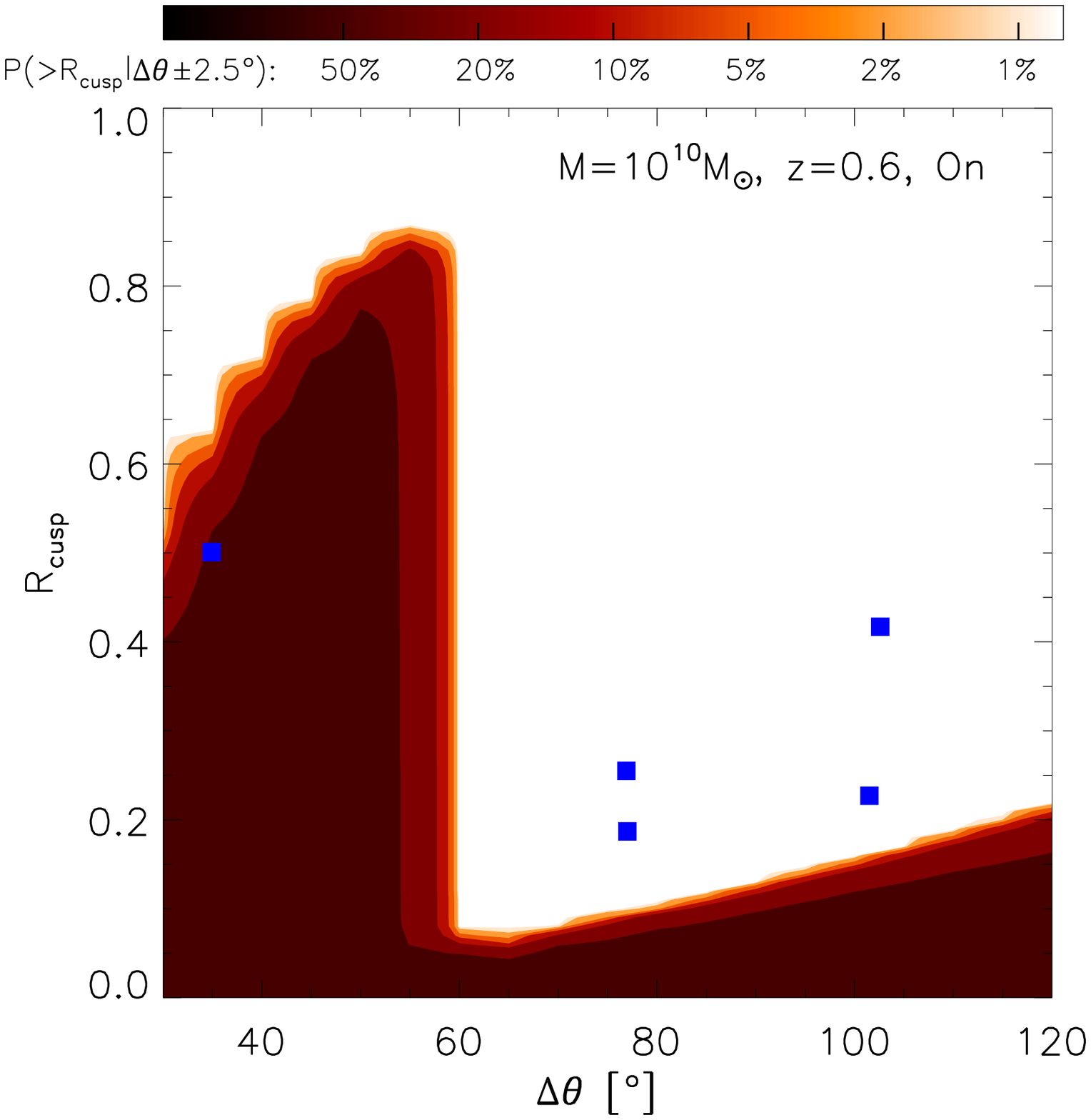}
\includegraphics[width=5.5cm]{count_M10Z0.6OutTSIS.eps}
\caption{Corresponding probability contour maps of $P(\geqslant \Rcusp
  | \Delta\theta \pm2.5^{\circ})$ for cases presented in
  Fig.~\ref{fig:CCZpos}. Symbols and contour levels are the same as in
  Fig.~\ref{fig:SIEandAq}. The top row presents cases where a
  perturber (of $10^{10} M_{\odot}$) is projected at the same angular
  position as shown in the upper panels of Fig.~\ref{fig:CCZpos}, but
  located at different redshifts: $z=0.4$ (foreground), $z=0.6$ (in
  the main-lens plane) and $z=1.4$ (background); the second row shows
  cases where the perturber is located at $z=0.6$ but projected at
  three different angular positions as shown in the lower panels of
  Fig.~\ref{fig:CCZpos}.}
\label{fig:ViolationZpos}
\end{figure*}

Fig.~\ref{fig:ViolationZpos} shows the contour maps of $P(\geqslant
\Rcusp | \Delta\theta \pm2.5^{\circ})$ in the $\Rcusp$-$\Delta\theta$
plane for the different scenarios above. Violation patterns as a
function of image opening angle $\Delta\theta$ vary with the positions
and redshifts of the perturbers. Notice that at redshifts greater than
that of the primary lens, the cone of light rays starts decreasing in
size towards the source. This means a ``background'' perturber that
appears to be projected close to the critical curve (where images
normally form) could actually be far away from the light ray. Such
perturbers would be less effective in causing convergence fluctuations
than their foreground counterparts (they would still contribute to the
shear field). However, depending on the distribution of the primary
lens and the source, there could be many more background structures
affecting the light ray than those in the foreground (see $\S$5).

\begin{figure*}
\begin{center}
\includegraphics[width=15cm]{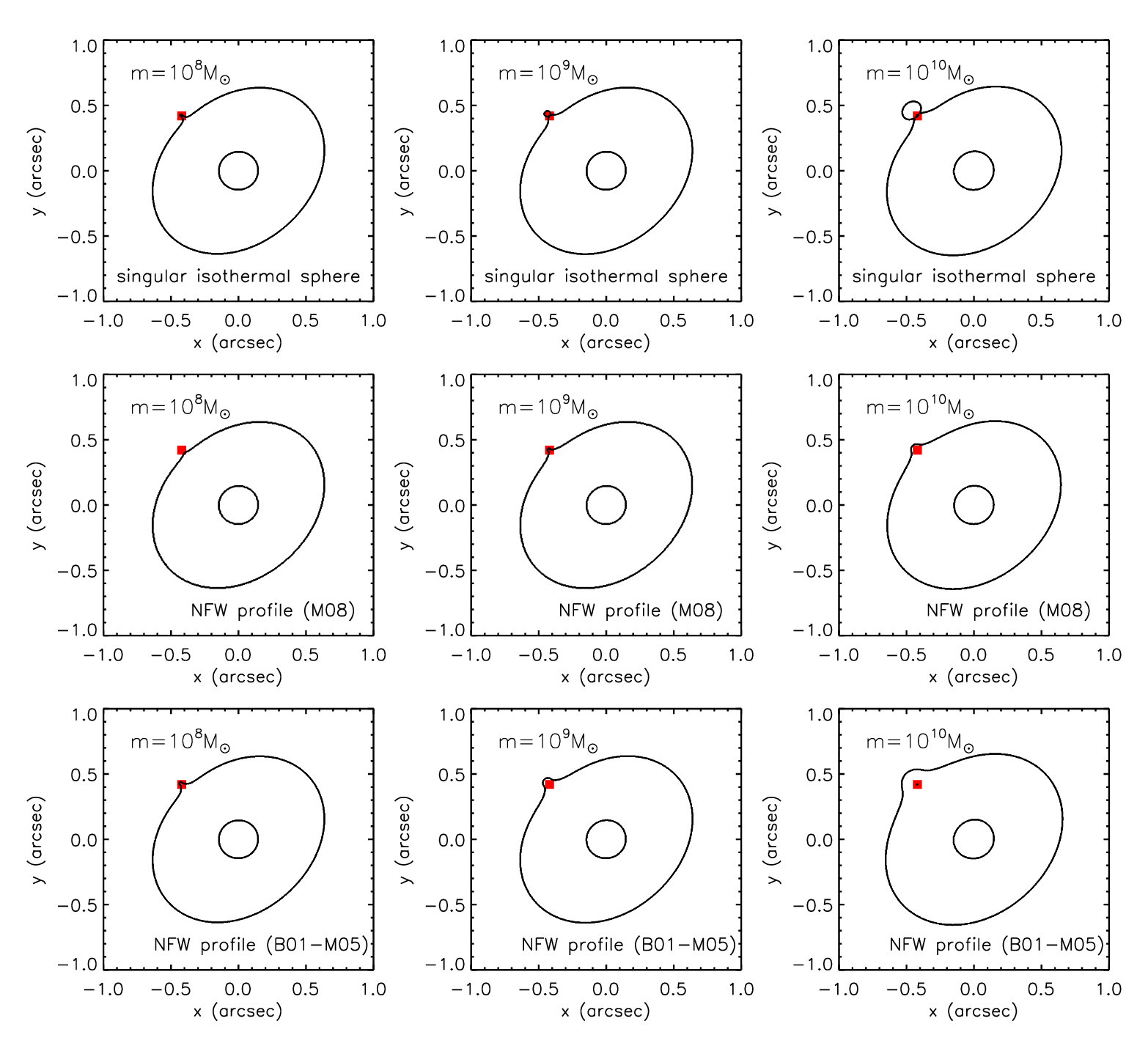}
\end{center}
\caption{Critical curves for smooth lens potential with a single
  perturbing halo (located at $z=0.6$), plotted as a red square. In
  columns from left to right the perturber has a mass of $10^{8}
  M_{\odot}$, $10^{9} M_{\odot}$, and $10^{10} M_{\odot}$,
  respectively. The rows from top to bottom correspond to different
  assumptions for the density profile: a truncated singular isothermal
  sphere, a truncated NFW profile with the M08 concentration-mass
  relation, and a truncated NFW profile with the B01-M05
  concentration-mass relation, respectively.}
\label{fig:CCprofiles}
\end{figure*}

The mass and density profile of a perturber also affect the production
of flux-ratio anomalies (and cusp-caustic violations) by altering the
effective cross-section. Singular isothermal spheres have been found
to be a good approximation for the inner density profiles of
relatively massive haloes
(\citealt{Rusin2003,TreuKoopmans04apj,RusinKochanek2005,Koopmans2006apj}),
where baryons are thought to dominate their central potentials. This
effect may be less important in smaller haloes, where the density
profile is more likely to be well approximated by the NFW distribution
characteristic of CDM haloes in $N$-body simulations
(\citealt{NFW97}). Nevertheless, there is still much controversy
whether observed low-mass haloes around dwarf galaxies have core-like
shallow profiles (\citealt{CorePIDwarfGalaxy2011}).

In the simple scenarios presented below and in our line-of-sight
lensing simulations (see $\S$5 and $\S$6), we model perturbing
haloes either as truncated singular isothermal spheres\footnote{A
  singular isothermal sphere may not be a realistic model for small
  haloes. We adopt this model for ease of comparison with previous
  work, e.g. \citet{Chen2003}.} or truncated NFW profiles, normalized
with their masses and truncated at their virial radii. We follow the
convention of defining the virial radius as $r_{200}$, the radius
within which the mean halo density is 200 times the critical density
of the Universe (at the appropriate redshift $z$). The mass enclosed
within $r_{200}$ is denoted as $M_{200}$. For the NFW profile, the
concentration parameter is $C_{200} \equiv r_{200}/r_{\rm s}$, where
$r_{\rm s}$ is the scale radius. This parameter is thought to
correlate with mass $M_{200}$ and redshift $z$. A number of
concentration-mass relations have been proposed in the literature,
based on $N$-body simulations.

In this work, we adopt the concentration-mass relation of
\citet{Maccio08CM} (hereafter M08) wherever we model perturbers as
truncated NFW profiles. The fitting formula (for a WMAP-1 cosmology,
close to that of the Millennium-II simulation) is given by:
\begin{equation}
  C_{200}(M_{200}, z) = \frac{10^{0.917}}{[H(z)/H_0]^{2/3}}
\bigg(\frac{M_{200}}{10^{12}M_{\odot}}\bigg)^{-0.104},
\label{Eq:NFWC_M08}
\end{equation}
where $H^2(z)=H_0^2 [\Omega_\Lambda+\Omega_m(1+z)^3]$.

The concentration-mass relation of \citet{BullockNFWC2001} was used by
\citet{Metcalf2005a,Metcalf2005b} to study how line-of-sight haloes
($10^{6}M_{\odot} \leqslant m \leqslant 10^{9}M_{\odot}$) contribute
to the flux anomaly problem. The adopted fitting formula was given by
(\citealt{Metcalf2005b}):
\begin{equation} C_{200}(M_{200}, z) = \frac{14}{1+z}
\bigg(\frac{M_{200}}{10^{12}M_{\odot}}\bigg)^{-0.15}.  \label{Eq:NFWC_B01}
\end{equation}
To compare with \citet{Metcalf2005a,Metcalf2005b}, we also perform our
analysis using this alternative concentration-mass relation,
hereafter referred to as B01-M05.

Fig.~\ref{fig:CCprofiles} shows how these different assumptions for
the mass, density profile and concentration-mass relation of a
perturber change the total critical curves produced by a primary lens
at $z=0.6$ and the perturbing halo located at the same redshift and
with a mass of $M_{200}=[10^8 M_{\odot},~ 10^9 M_{\odot},~ 10^{10}
M_{\odot}]$. A different density profile (a truncated singular
isothermal sphere, a truncated NFW profile with the M08
concentration-mass relation and a truncated NFW profile with the
B01-M05 concentration-mass relation) is assumed in each row of
Fig.~\ref{fig:CCprofiles}. Different distortions to the critical
curve correspond to different levels of violations in the smooth-lens
flux-ratio relationship.

The mass dependence of the violation pattern has been studied
systematically, with results presented in $\S$6, which also includes a
discussion of effects from different concentration-mass relations and
from allowing scatter in the concentration on the overall
cusp-violation probabilities.

In this section we have illustrated the effects of varying the
redshift, impact parameter, mass and density profile of a single
perturbing halo. In practice, perturbations could arise anywhere along
the line of sight and from many different haloes. The overall
perturbation is far more complicated than any of the simple cases
presented here. In the following sections, we use cosmological
$N$-body simulations to obtain self-consistent and realistic
distributions of perturbers along strong lensing sight lines, and
estimate the net perturbation and the likelihood of the observed flux
ratio violations.

\section{Lensing Lines of Sight from Cosmological Simulations}

\subsection{Constructing lensing cones from MS-II}

The Millennium-II simulation (MS-II; \citealt{Millennium2}) is an
$N$-body simulation of a cubic cosmological volume with a comoving
side length of 100$h^{-1}~\Mpc$, at a spatial resolution of
1$h^{-1}~\kpc$ and mass resolution of $6.89\times10^6
h^{-1}M_{\odot}$. The cosmological parameters of MS-II are the same as
those of the earlier Millennium and Aquarius simulations, consistent
with the WMAP-1 results. MS-II provides us with the large-scale
distributions of a cosmological sample of dark matter haloes. When
tracing lensing sight lines through this simulation, we use the
following method to determine where haloes cross the past light cone of
a fiducial observer (for more details, see \citealt{Raul2008thesis}).

\begin{figure}
\begin{center}
\includegraphics[width=8cm]{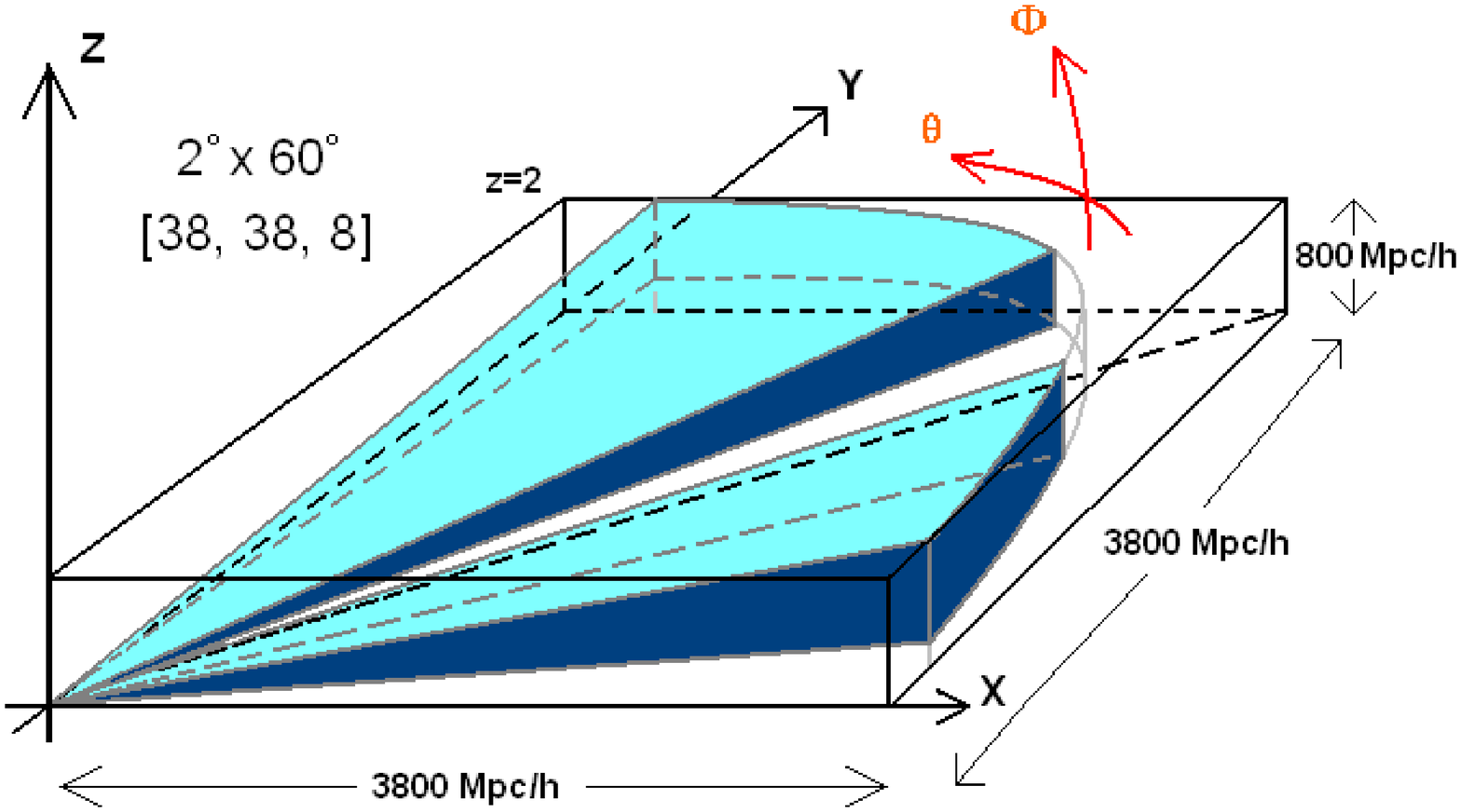}
\end{center}
\caption{The geometry of the replicated box for light-cone generation:
  the MS-II simulation box of 100$h^{-1}$ Mpc is repeated in its $X$,
  $Y$ and $Z$ dimensions as many times as needed to cover the desired
  redshift range and angular size. For a source redshift of $z_s=2$,
  the total dimension of the combined box is set to be
  $38\times38\times8$, in units of one Millennium-II simulation box. An
  observer is put at the origin (0, 0, 0) of this box. The
  position angle pair $(\theta, \phi)$ of a given line-of-sight vector
  are defined as the angles measured from the $ZX$- and $XY$-plane,
  respectively. The simulated sky we have looked at is then two
  $2^{\circ} \times 30^{\circ}$ stripes, which cover $10^{\circ}
  \leqslant \theta \leqslant 40^{\circ}$ and $50^{\circ} \leqslant
  \theta \leqslant 80^{\circ}$, and $10^{\circ} \leqslant \phi
  \leqslant 12^{\circ}$. Directions along the axes (with $\leqslant
  10^{\circ}$) and along $\theta\sim45^{\circ}$ have been excluded to
  avoid significant structure repetition. } \label{fig:RepBoxMSII}
\end{figure}

We start by replicating the 100$h^{-1}$~Mpc simulation box in its $X$,
$Y$ and $Z$ dimensions, as many times as we need to cover the desired
redshift range (along the sight line) and angular size (transverse to
the sight line). For computational efficiency we only let the combined
box go to the source redshift in the $X$ and $Y$ dimensions, and keep
the number of replications in the $Z$ dimension to a minimum. As
illustrated in Fig. \ref{fig:RepBoxMSII}, the observer is located at
the origin (0, 0, 0) in one corner of this replicated box. Assuming a
source redshift of $z_s=2$, the total dimension of the combined box is
chosen to be $38\times38\times8$, in units of one Millennium-II
simulation box. The position angles $(\theta, \phi)$ of a given
line-of-sight vector are defined as the angles measured from the $ZX$-
and $XY$-plane, respectively. The simulated sky into which we trace
sight lines is then two $2^{\circ} \times 30^{\circ}$ stripes, which
cover $10^{\circ} \leqslant \theta \leqslant 40^{\circ}$ and
$50^{\circ} \leqslant \theta \leqslant 80^{\circ}$, and $10^{\circ}
\leqslant \phi \leqslant 12^{\circ}$. Directions along the $X$ and $Y$
axes (with $\leqslant 10^{\circ}$) and along $40^{\circ} \leqslant
\theta \leqslant 50^{\circ}$ have been excluded to avoid significant
repetition of structures in the replicated box.

Haloes at each simulation snapshot (corresponding to a particular
redshift) are identified using the Friends-of-Friends algorithm
(\citealt{Davis1985FoF}). Haloes also contain many subhaloes; these
are identified using the {\sc subfind} algorithm
(\citealt{VolkerGADGET2}). The minimum mass of subhaloes resolved by
the simulation is $1.4\times10^8 h^{-1}M_{\odot}$ (corresponding to 20
particles). Haloes at different snapshots are linked together by an
algorithm for defining their merging history (\citealt{Helly03}). We
follow haloes in each of these merger trees and predict their
trajectories (in the replicated box) between every two adjacent
snapshots. In this way, we can find the exact redshift and comoving
position of a halo at the moment it crosses the past light cone of the
observer. When a halo crosses the light cone, all its subhaloes are
assumed to cross at the same redshift. We assume that the relative
positions of these subhaloes at the light-cone crossing time are the
same as in the previous snapshot. \\

Hereafter, we will use the term ``lensing cone'' to refer to the
observer's light cone that encloses a particular lensing sight line
towards a certain direction in the sky (and out to the source
redshift). All haloes that cross the past light cone are checked to
see if they are physically within a given lensing cone. If so, their
positions, redshifts, masses and half mass radii are stored for lens
modelling. All lensing cones are $50\arcsec\times 50\arcsec$-wide, out
to a source redshift $z_s=2$, and contain a primary lens around
redshift $z_d=0.6$ (typical source and lens redshifts for the observed
quasar lensing systems).

To build up our lensing cone catalogue, we randomly select about 300
directions in the $2^{\circ} \times 60^{\circ}$ simulated sky, each of
which goes through at least one galaxy-scale halo with a mass above
$10^{12}h^{-1}M_{\odot}$ located at redshift $|z-0.6| \leqslant 0.02$
in the replicated box. This ensures that the primary haloes we select
are responsible for producing multiple images of the $z_s=2$
background sources. We have confirmed that these $\sim$ 300 randomly
selected primary lenses are representative of mass and circular
velocity distributions of $\sim$23,000 haloes that meet the same
selection criteria in the simulated sky. 

As stated in \S4.2, velocity dispersions of the simulated main lensing
haloes range from 200$\kms$ to 300$\kms$, comparable to our sample of
observed lenses (see the end of \S3.1). As the selection function of
the observed lenses is hard to define, we assume that they are a
random sample of haloes in the ranges of velocity dispersion and
redshift given above. Furthermore, by imposing a lower mass limit of
$10^{12}h^{-1}M_{\odot}$ on the main lens, we have excluded cases
where two or more less massive haloes aligned along the same line of
sight produce a comparable strong-lensing signal. However, the chance
is small for two foreground haloes, both more massive than
$10^{10}h^{-1}M_{\odot}$, to be well aligned to jointly lens a
background galaxy.
  Emprically, such ``three-dimensional'' lenses are rare: in the CLASS
  survey, only one probable such case has been reported out of a total
  of 22 candidates (\citealt{CMA01B2114,Augusto01B2114}).

\begin{figure} \includegraphics[width=8cm]{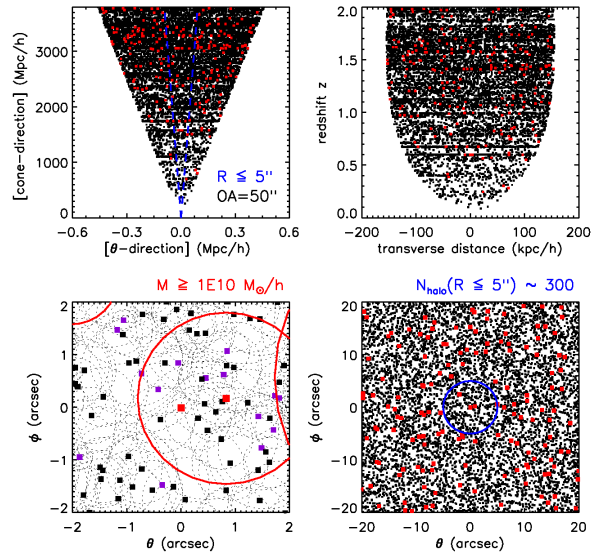}
  \caption{The halo distribution within an example lensing cone in a
    slice of depth 3800$h^{-1}$ Mpc (in co-moving distance), out to a
    redshift of $z=2$. The cone covers a region of $50\arcsec\times
    50\arcsec$. In all four panels, haloes and subhaloes are plotted
    as black squares; those more massive than $10^{10} M_{\odot}$ are
    shown by red squares. The central region of radius $R\leqslant
    5\arcsec$ is indicated with blue lines in both top left and bottom
    right panels. Top left: the projected cone geometry in comoving
    distance. Top right: same lensing cone, shown in the
    $redshift-physical~distance$ plane. Bottom left: the central
    $4\arcsec\times 4\arcsec$ region of the light cone projected in
    the sky, where the main tangential critical curves form; subhaloes
    are indicated by purple squares in this panel, circles in solid
    red and dashed black indicate the half-mass radii of haloes (and
    subhaloes). Bottom right: an expanded view, showing haloes and
    subhaloes projected within the central $40\arcsec\times 40\arcsec$
    of the light cone. }
\label{fig:LCOA50} \end{figure}

Fig.~\ref{fig:LCOA50} shows the geometry and halo distribution of an
example lensing cone. All haloes within a given cone are used for the
lensing calculation. On average each lensing cone (of $50\arcsec\times
50\arcsec$) contains about $10,000$ ($12,000$) haloes
(subhaloes). Within a projected central region of $R\leqslant
5\arcsec$ for strong lensing, there are on average $\sim300$ haloes
with $m>10^8h^{-1}M_{\odot}$ directly contributing to the convergence
field. The rest are distributed further out (in projection) and
contribute to the shear field of this region in the same way as point
masses.

\subsection{Ray-tracing through MS-II lensing cones ~~~~~~~~~~~~~~and line-of-sight lens modelling}

To carry out calculations for multi-plane light deflection, we assume
60 lens planes distributed with equal spacing in redshift between the
observer and the source at $z_s=2$. In each of these lens planes, a
region of $5\arcsec\times5\arcsec$ around the line centre is covered
by a $1000\times1000$ rigid grid in order to calculate the Jacobian
matrix $A_{\rm s}$ (Eq.~\ref{eq:Jacobian}) between the source plane and the
final image plane.

Haloes within a lensing cone are projected into these lens planes
according to their redshifts. The main lens halo is modelled as an
isothermal ellipsoid, for which a universal axis ratio ($q_3=0.8$) and
core radius ($s_0=0.05\arcsec$) are assumed. The orientation of the
ellipsoid is randomly chosen in the interval of [0, 2$\pi$]. The
lensing strength $b_{\rm SIE}$ (related with $b_I$ through $b_I=b_{\rm
  SIE}e/sin^{-1}e$, where $e=(1-q_3^2)^{1/2}$, see
\citealt{KKprofile1998}) is derived through an empirical relationship
between halo's virial velocity $V_{200}$ and the velocity dispersion
$\sigma_{\rm SIE}$ of the equivalent isothermal ellipsoid
(\citealt{ChaeMao2006}):
\begin{equation} \frac{\sigma_{\rm SIE}}{200~{\rm km/s}} \approx
  \bigg[\frac{1.17~V_{200}}{200~{\rm km/s}} \bigg]^{0.22} ~ (171~{\rm
    km/s} \leqslant V_{200} \leqslant 563~{\rm
    km/s}) \label{eq:V200SigmaSIE}
\end{equation} 
and $b_{\rm SIE}=4\pi(\sigma_{\rm SIE}/c)^2D_{\rm ds}/D_{\rm s}$,
where $c$ is the speed of light and $D_{\rm ds}$ and $D_{\rm s}$ are
the angular diameter distances between the main lens and the source,
and the source and the observer, respectively. The virial velocity
$V_{200}$ is obtained from halo mass $M_{200}$ and its virial radius
$r_{200}$ through $V_{200}^2=GM_{200}/r_{200}$. Our requirement that
the main lens be more massive than $10^{12} h^{-1}M_{\odot}$ results
in $\sigma_{\rm SIE}$ ranging from 200$\kms$ to 300$\kms$, with a
weighted mean (by cross-section, $\propto \sigma^4$) of 222$\kms$
corresponding to a lensing strength of $b_{\rm SIE}=0.84\arcsec$ for
our adopted lens and source redshifts ($z_d=0.6, z_s=2.0$).

Within each lensing cone, haloes with projected profiles that are
completely outside the central $5\arcsec\times5\arcsec$ region are
treated as point masses. Those within this region are assigned a
density profile: as described above, we investigate three distinct
choices of this profile (singular isothermal sphere, NFW with the M08
concentration-mass relation, and NFW with the B01-M05
concentration-mass relation). All halo profiles are normalized to
their masses $M_{200}$ and truncated at the virial radii $r_{200}$;
subhaloes are truncated at two times their half mass radii.

For each line of sight, deflection angles are individually calculated
for the equivalent isothermal ellipsoid of the main lens and for all
line-of-sight (sub)haloes, and are tabulated to the meshes at
different lens planes. Through ray tracing, source positions
$\vec{\beta}_N$ that correspond to the final image plane
$\vec{\theta}_1$ are identified, and the final Jacobian matrix
$A_{\rm s}={\partial{\vec{\beta}_N}}/{\partial{\vec{\theta}_1}}$ is then
derived using the finite differencing method. Images of a given source
position are effectively found using the Newton-Raphson iteration
method.

\section{Results from the Millennium II Simulation}

In order to calculate $P(\geqslant \Rcusp | \Delta\theta
\pm2.5^{\circ})$ -- the probability distribution in the
$\Rcusp-\Delta\theta$ plane for individual lensing cones, we generate
$10,000\sim20,000$ cusp sources whose close triple images have image
opening angles $\Delta\theta \leqslant 120^{\circ}$. We have also
calculated $P^{90}(\Rcusp^{0.187})$ for cases with $\Delta\theta
\leqslant 90^{\circ}$ as an overall estimate for cusp-caustic
violations to compare with our previous work, in which only cases with
$\Delta\theta \leqslant 90^{\circ}$ were examined for violations
(caused by intrinsic substructures within the main lens).

To derive average violation probabilities over all sight lines, we
weight $P(\geqslant \Rcusp | \Delta\theta \pm2.5^{\circ})$ and
$P^{90}(\Rcusp^{0.187})$ of each lensing cone by the quadruple-image
lensing cross-section in the source plane (simply the fractional area
within the tangential caustic). We do not account for magnification
bias among the cusp sources. 

\begin{table*}
\centering \caption{$P^{90}(\Rcusp^{0.187})$ from cosmological
simulations.} \small\addtolength{\tabcolsep}{-1.5pt}
\label{tab:MSIIlosRcusp}
\begin{minipage} {\textwidth}
\begin{tabular}[b]{c|c|c|c}\hline
  line-of-sight perturbers & truncated singular isothermal sphere & truncated NFW (M08) & truncated NFW (B01-M05) \\
  \hline 
  \multicolumn{1}{l}{~~~haloes + subhaloes} & 7.8\% & 7.3\% & 12.8\% \\
  \multicolumn{1}{l}{~~~haloes only}  & 7.0\% & 5.2\% & 9.7\% \\
  \multicolumn{1}{l}{~~~background haloes}  & 5.0\% & 3.2\% & 5.8\% \\
  \multicolumn{1}{l}{~~~foreground haloes}  & 2.0\% & 1.9\% & 4.0\% \\
  \hline
\end{tabular}
\end{minipage}
\begin{flushleft}
  Note: $P^{90}(\Rcusp^{0.187})\approx 10\%$ was derived using only
  substructure populations ($m\gtrsim 10^5 h^{-1}M_{\odot}$) from the
  Aquarius simulations (\citealt{Dandan09AquI,Dandan2010AqII}). Cases
  here are using line-of-sight structures (haloes and subhaloes) from
  the Millennium II simulation ($m > 10^8 h^{-1}M_{\odot}$); subhaloes
  from the main lensing halo have been excluded. \\
\end{flushleft}
\end{table*}

\begin{figure*} \includegraphics[width=5.5cm]{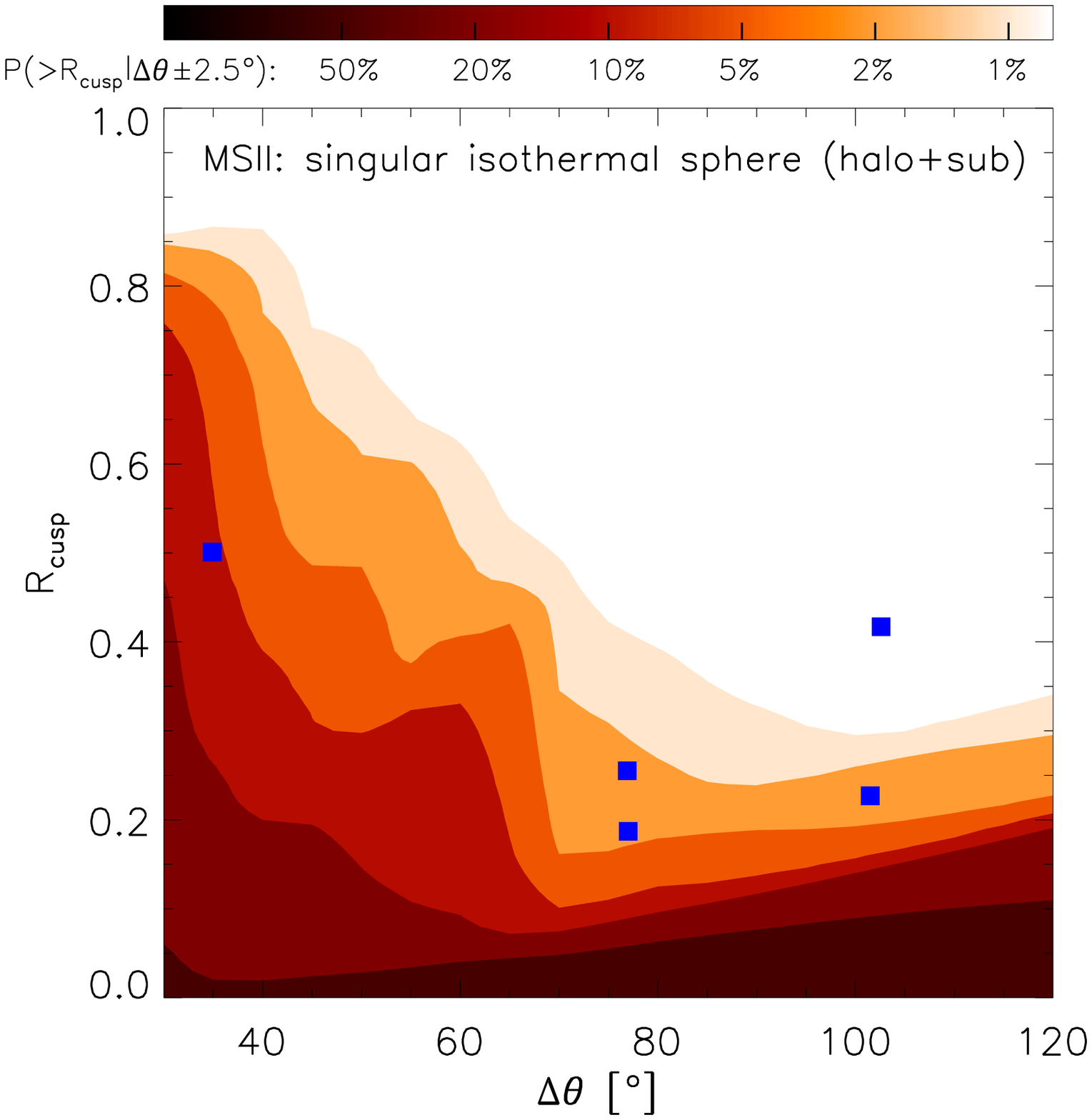}
\includegraphics[width=5.5cm]{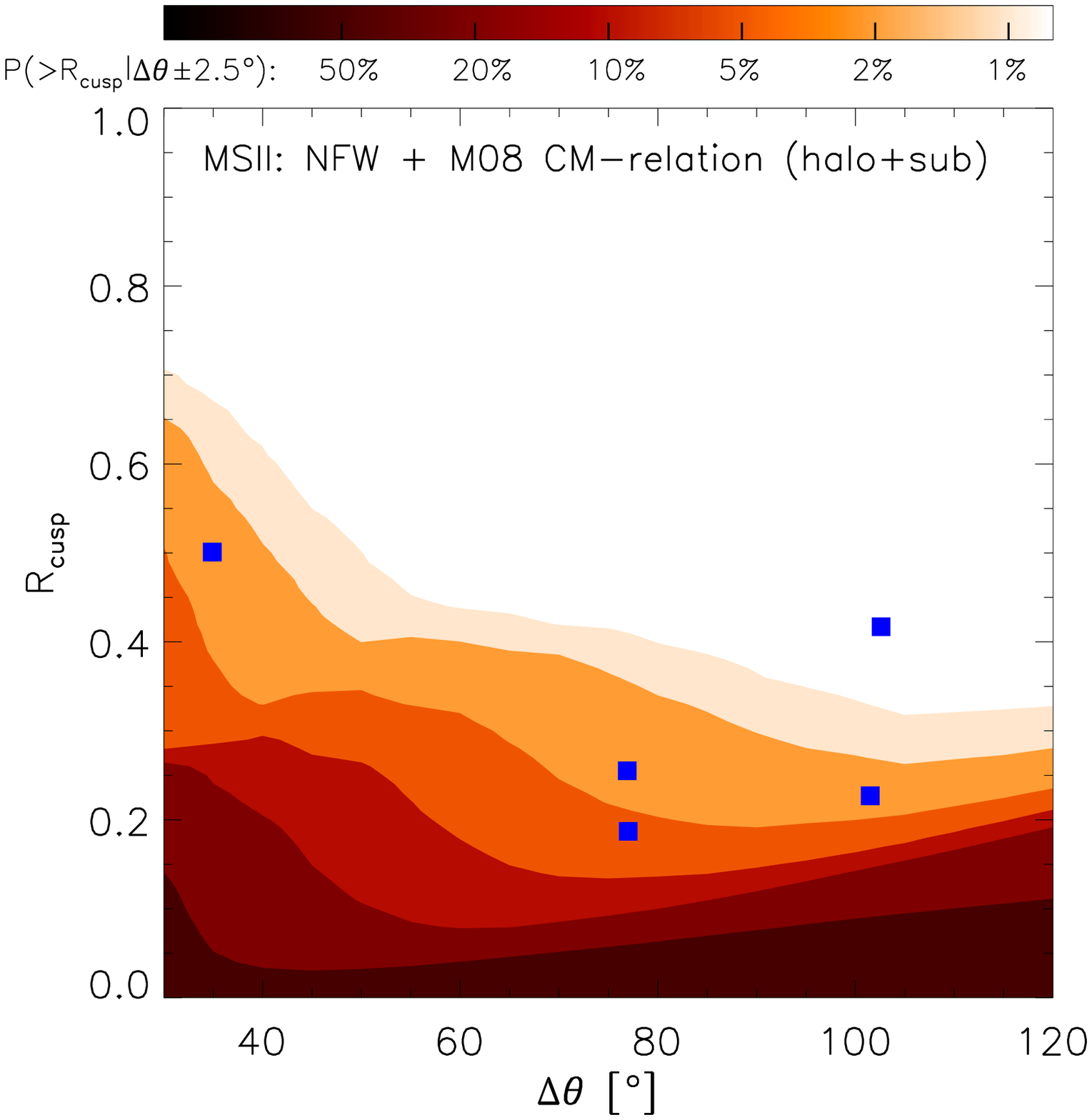}
\includegraphics[width=5.5cm]{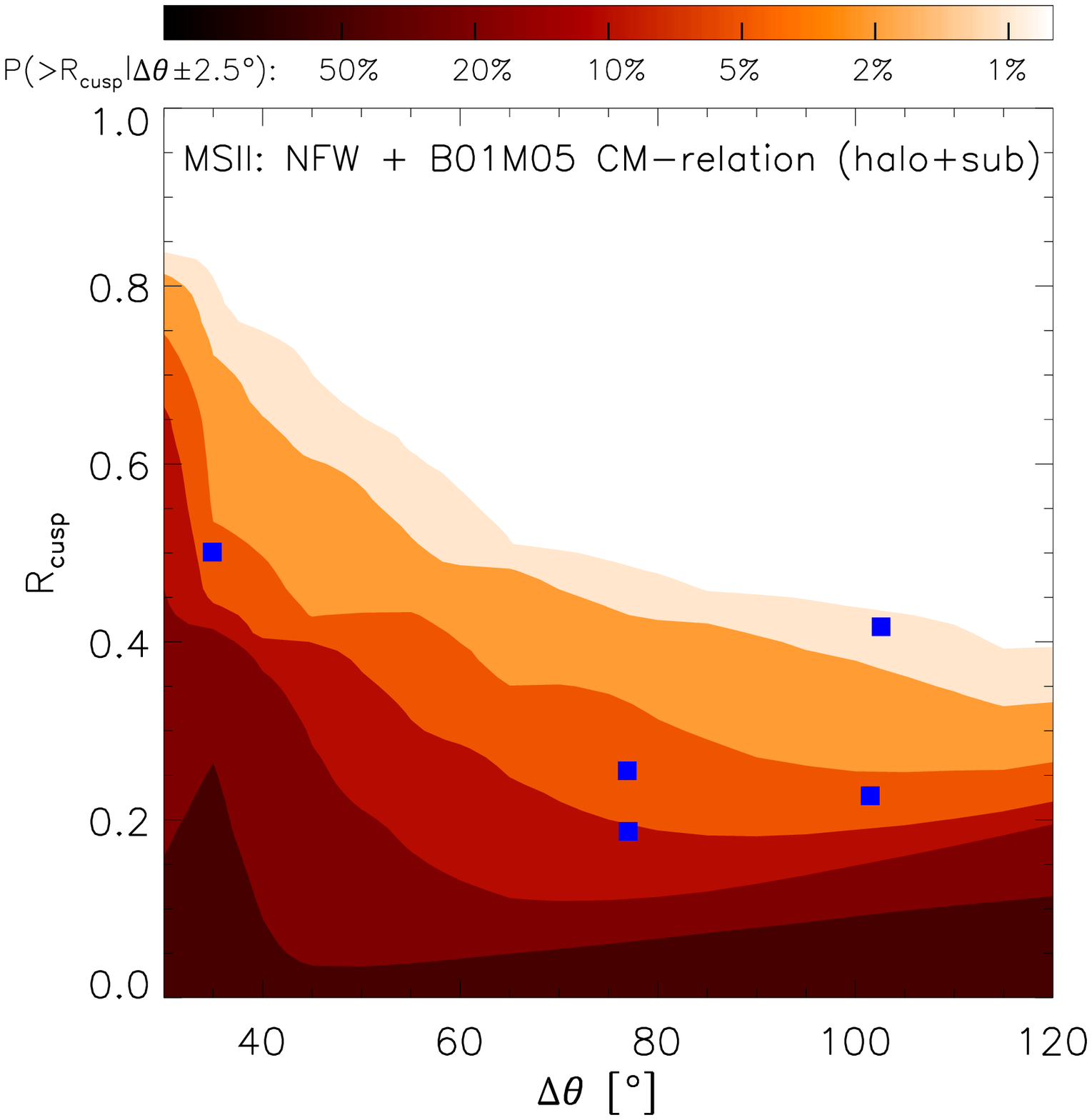} \\
\includegraphics[width=5.5cm]{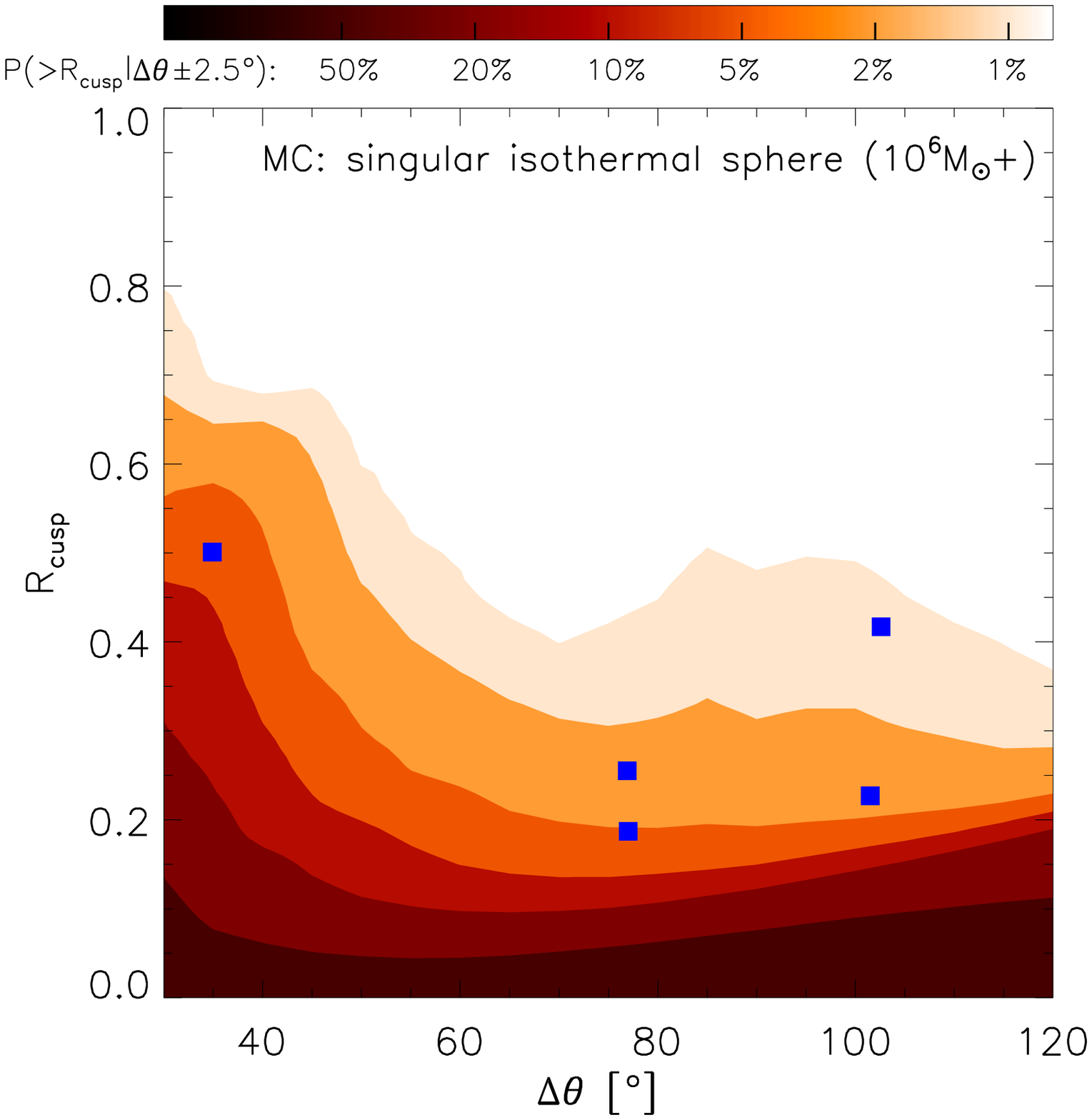}
\includegraphics[width=5.5cm]{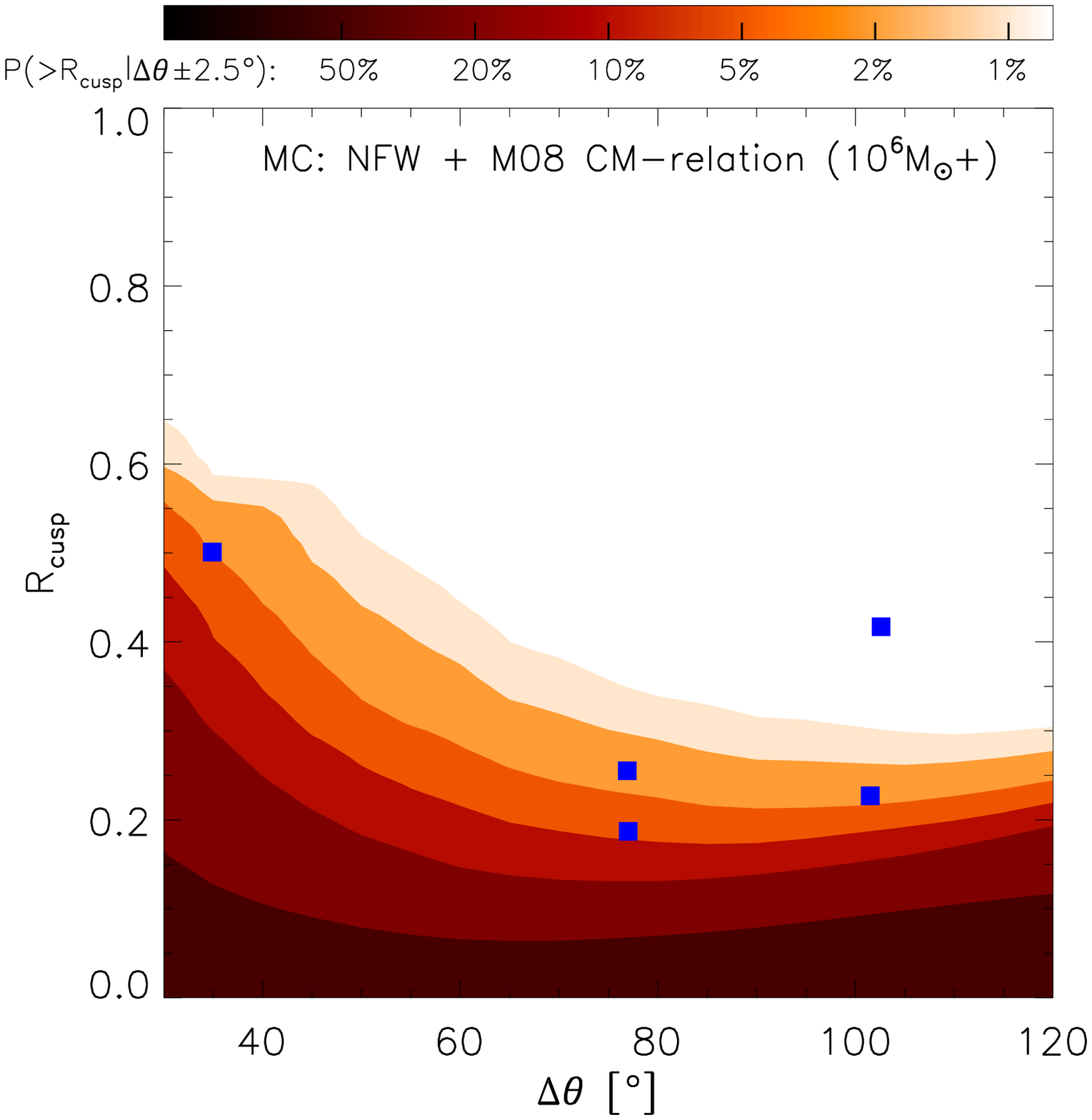}
\includegraphics[width=5.5cm]{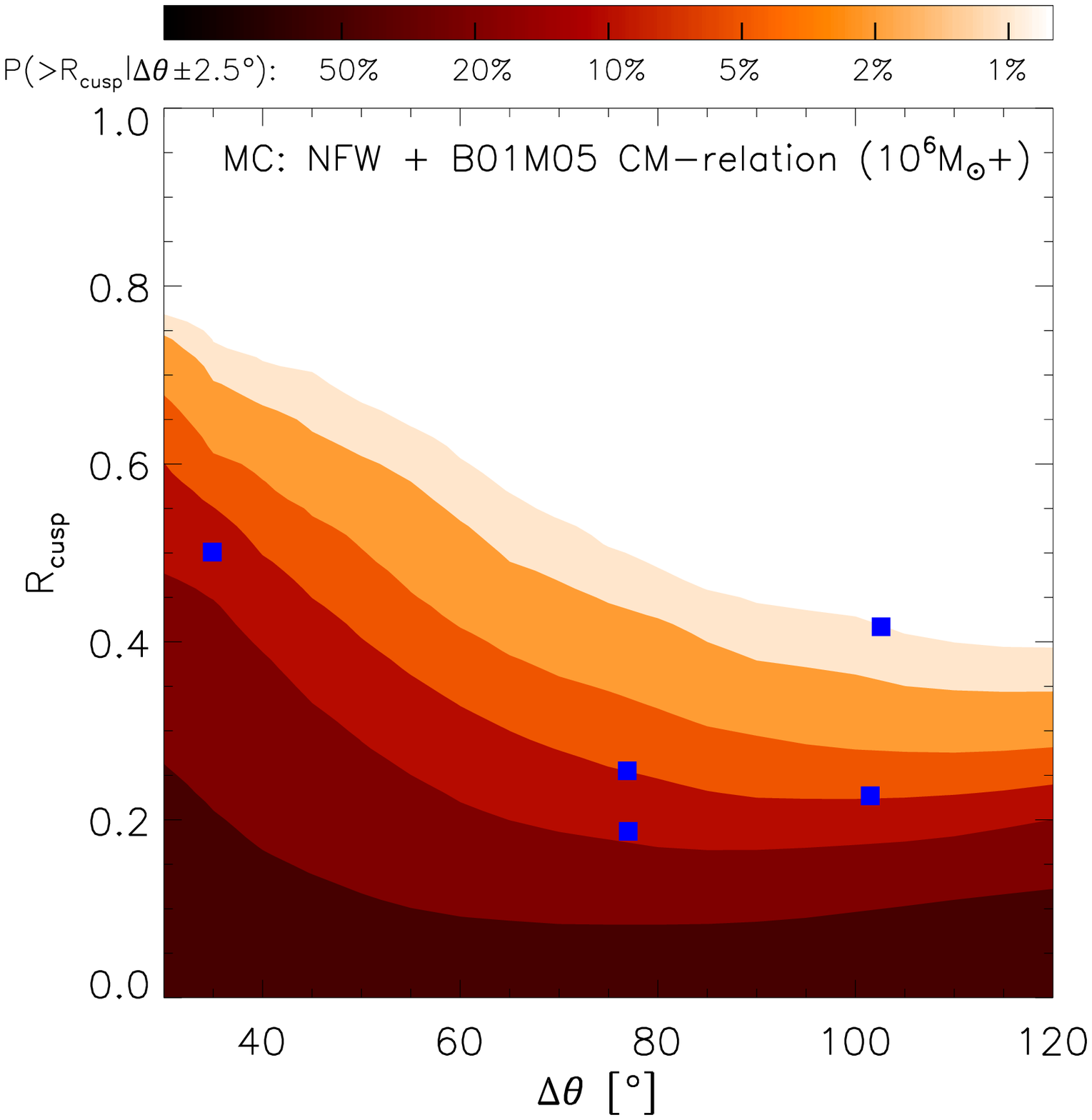} \caption{Contour
  maps of the violation probability $P(\geqslant \Rcusp | \Delta\theta
  \pm2.5^{\circ})$.  Symbols and contour levels are the same as in
  Fig.~\ref{fig:SIEandAq}. The upper panels are for MS-II lensing
  cones with all line-of-sight haloes and subhaloes ($m > 10^8
  h^{-1}M_{\odot}$), and the bottom panels are for Monte-Carlo lensing
  cones with line-of-sight haloes that follow the Sheth-Tormen mass
  function ($m \geqslant 10^6 h^{-1}M_{\odot}$). Three columns from
  left to right correspond to different assumptions for the density
  profile: truncated singular isothermal spheres, truncated NFW
  profiles with M08 and B01-M05 concentration-mass relations,
  respectively. } \label{fig:losvsmainSISNFW}
\end{figure*}

Fig.~\ref{fig:losvsmainSISNFW} (upper panels) and Table
\ref{tab:MSIIlosRcusp} show that if perturbing structures are haloes
(and subhaloes) distributed outside the main lens along the line of
sight as in the Millennium-II simulation, where such haloes are
resolved to $10^8 h^{-1}M_{\odot}$, they cause a non-negligible amount
of cusp violations, comparable to those due to the substructures in
the lens itself. However, the violation pattern (as a function of
$\Delta\theta$) depends strongly on the density profiles applied to
haloes projected near the centre of the line of sight.

\subsection{Effects from massive line-of-sight haloes}

\begin{figure*}
\includegraphics[scale=0.3]{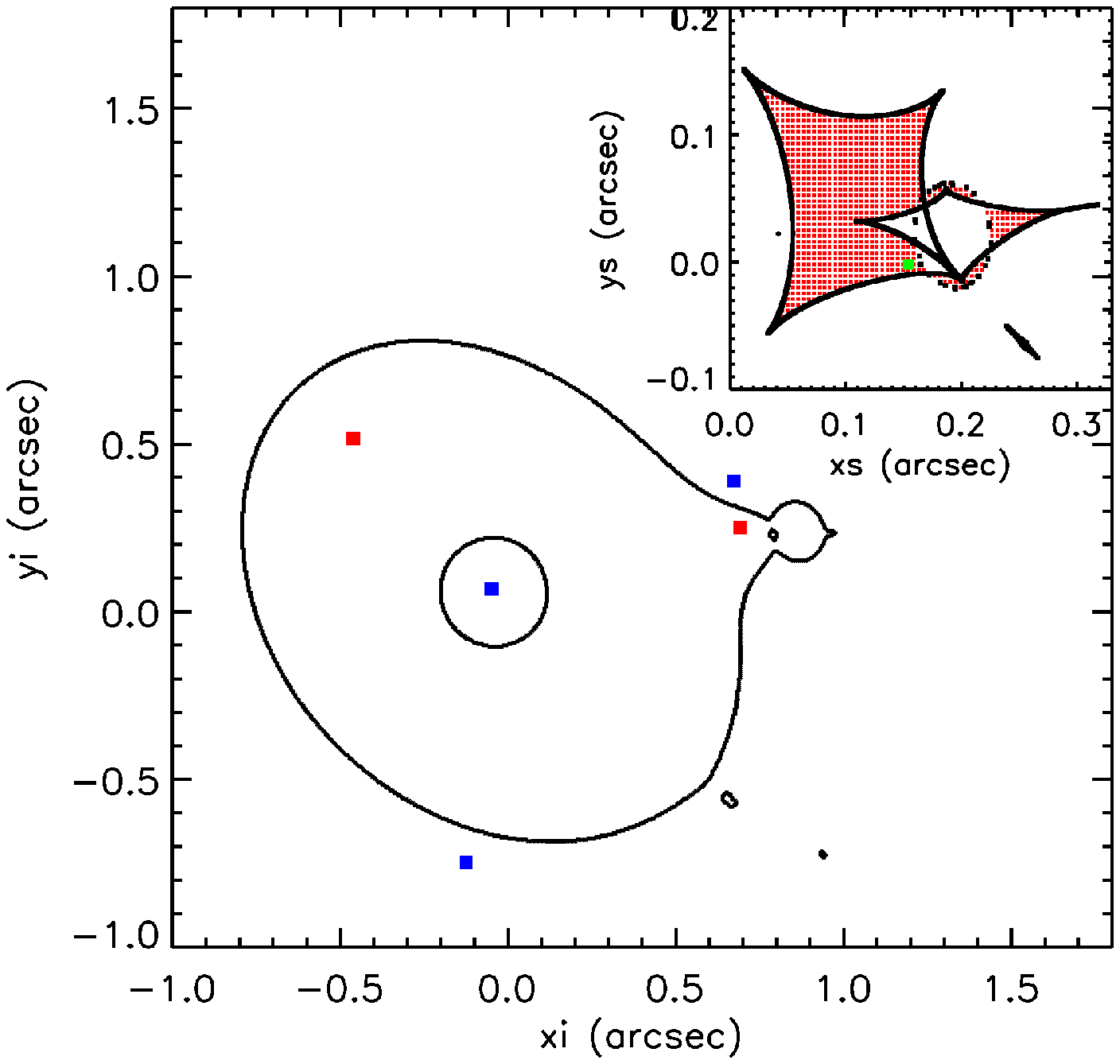}
\includegraphics[scale=0.3]{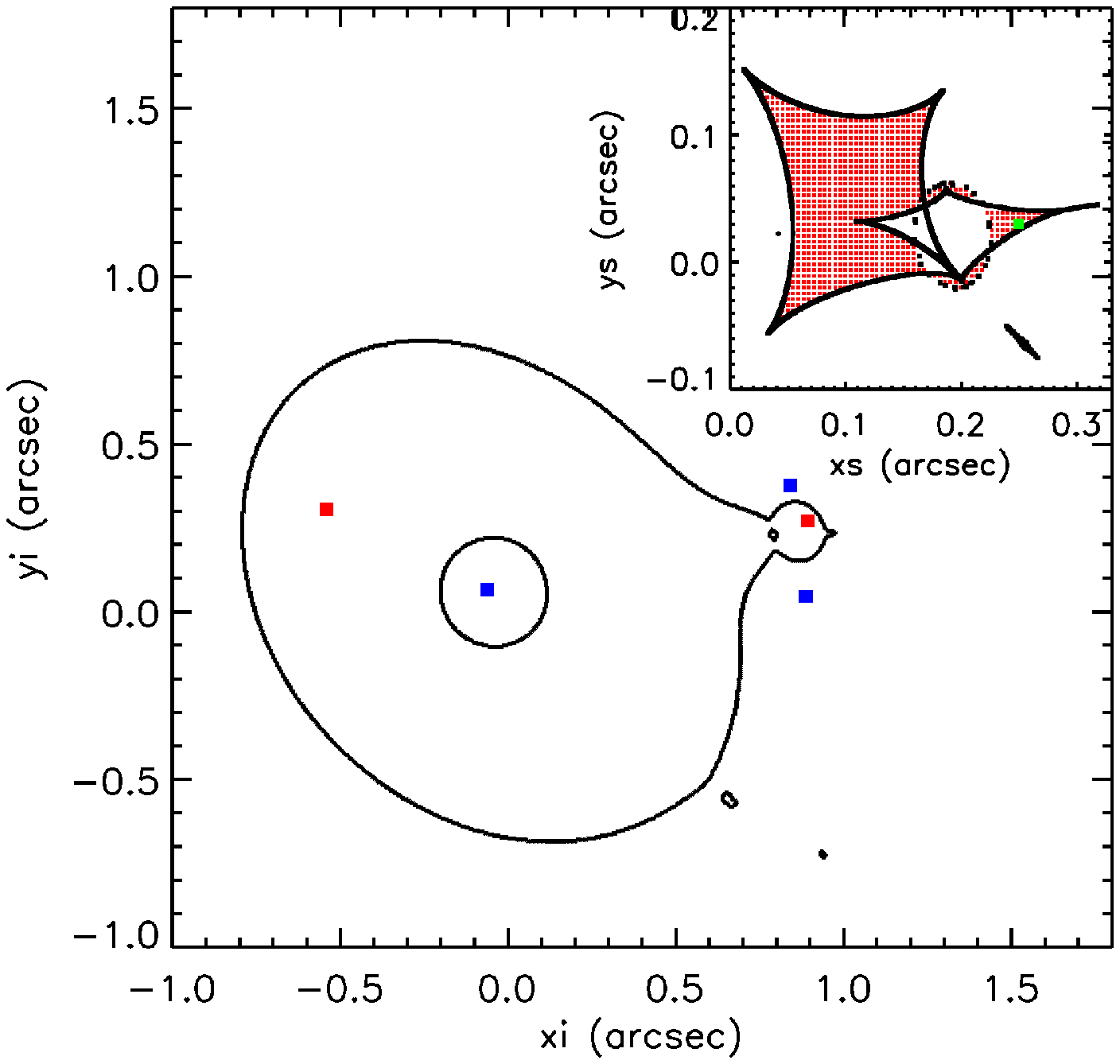}
\includegraphics[scale=0.3]{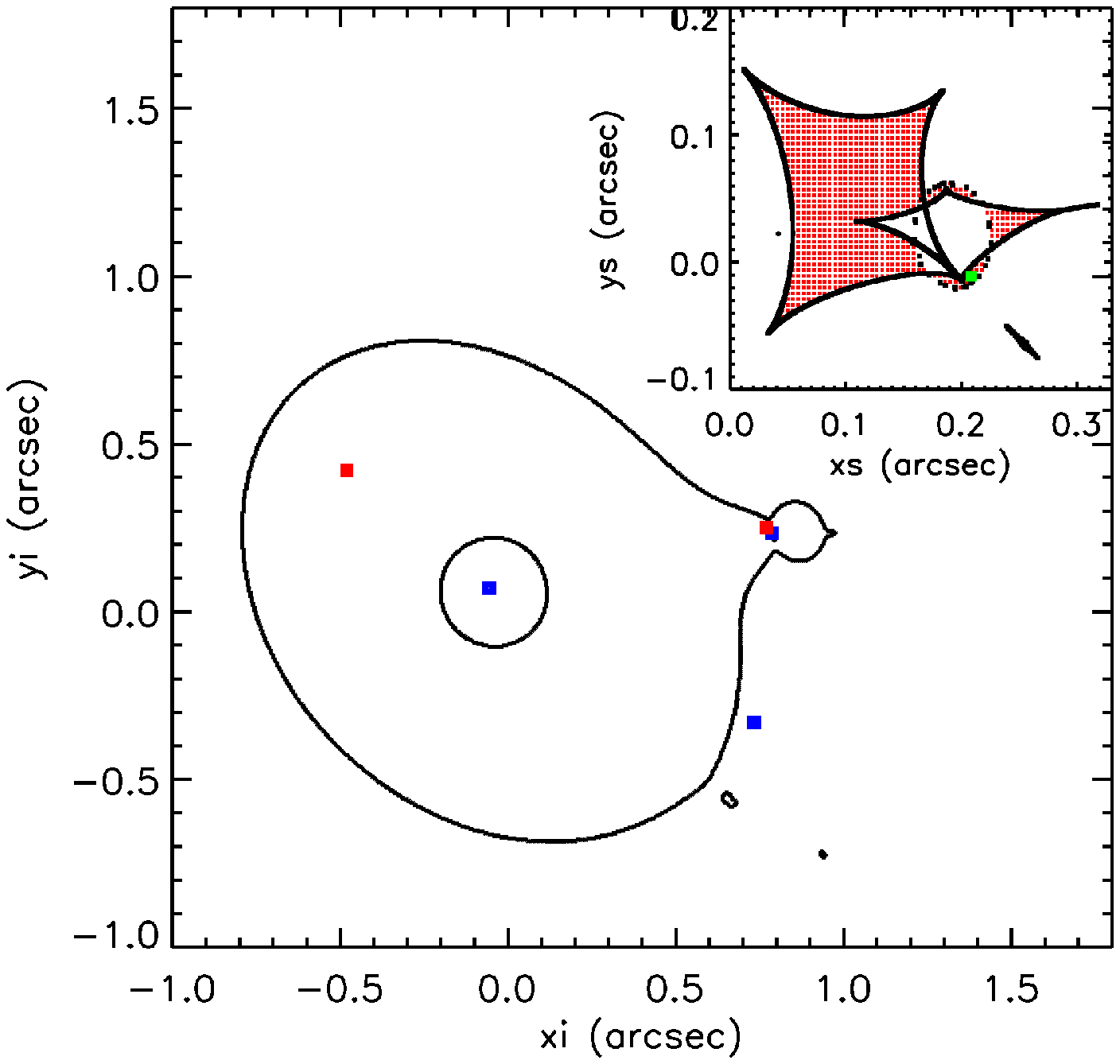}
\includegraphics[scale=0.3]{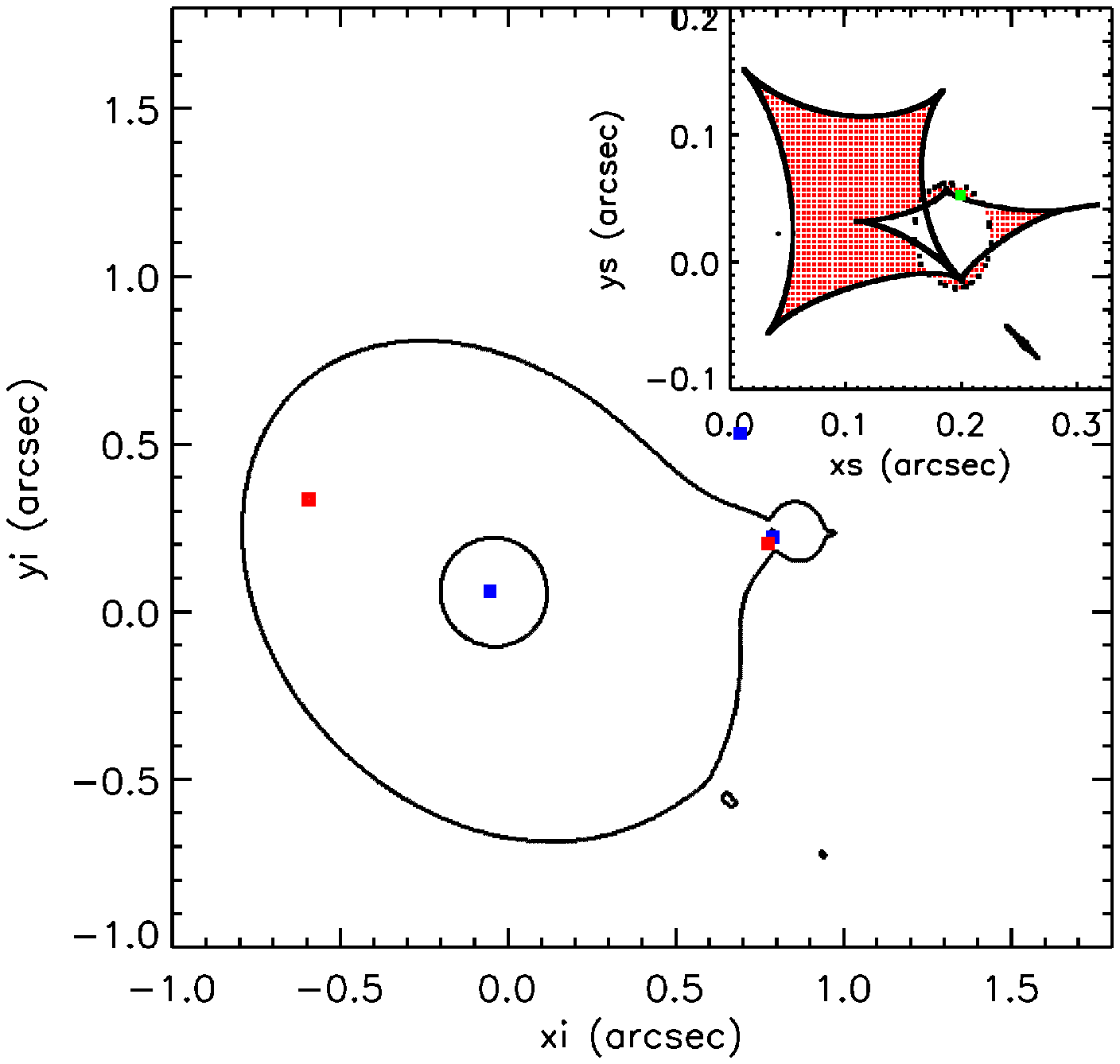}
\includegraphics[scale=0.3]{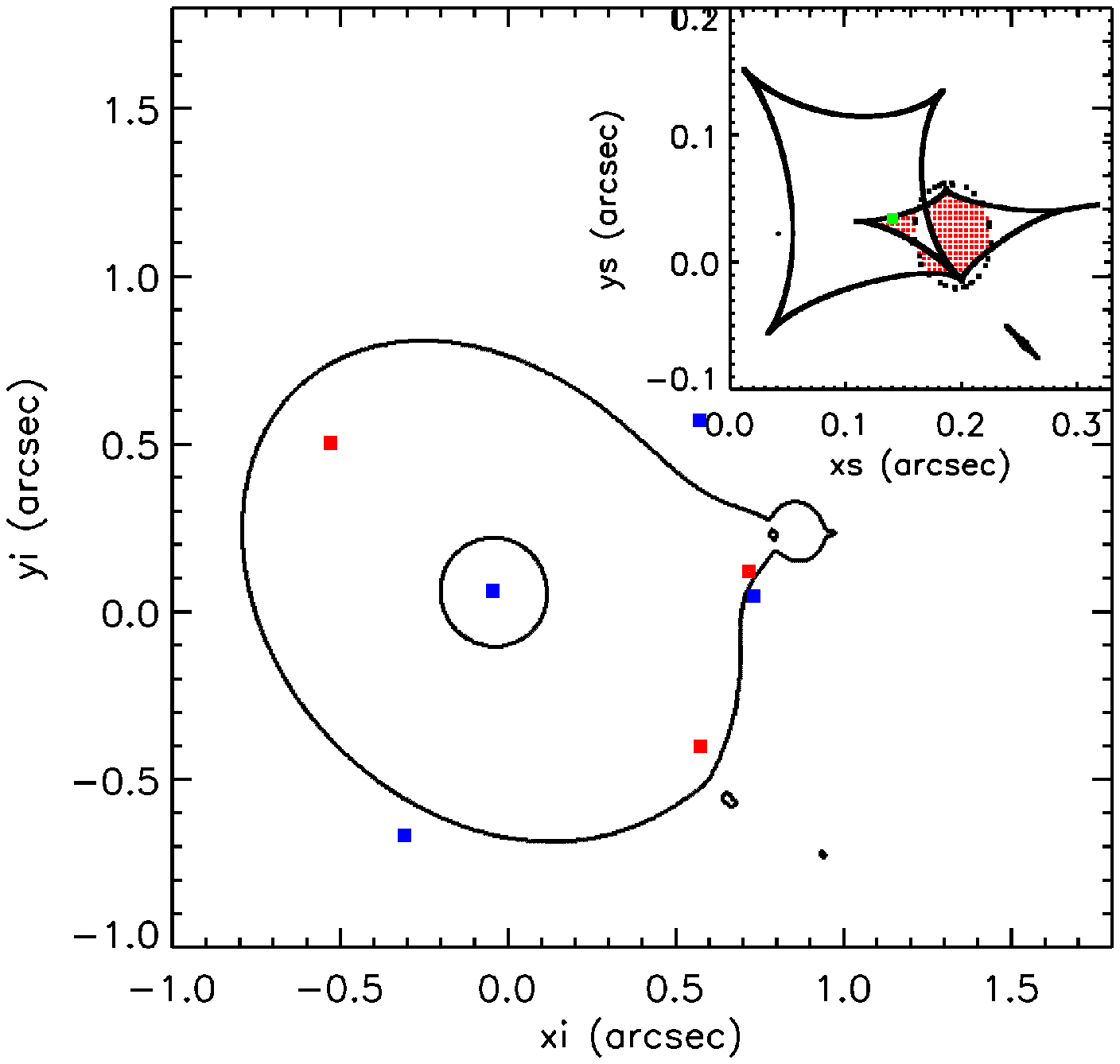}
\includegraphics[scale=0.3]{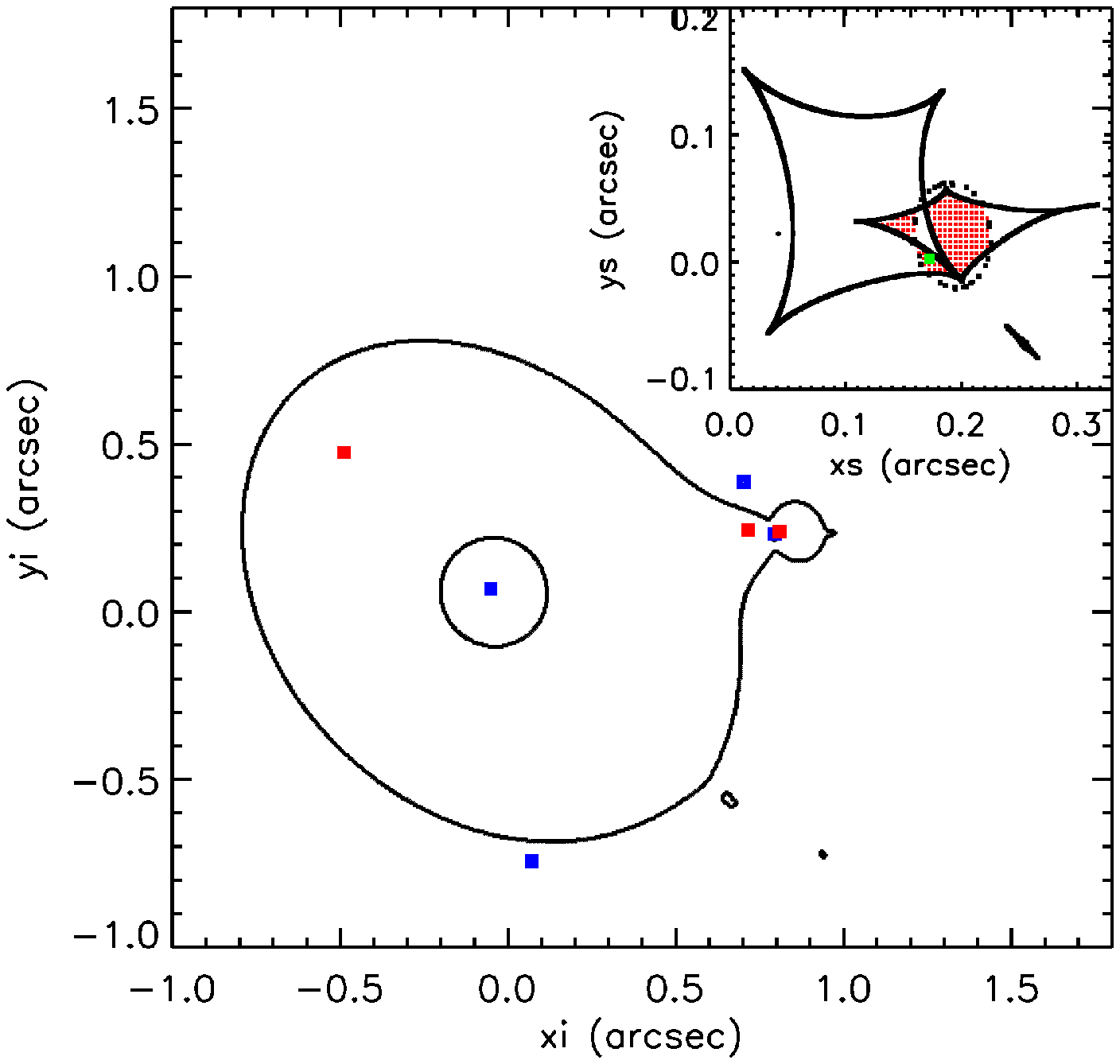}
\includegraphics[scale=0.3]{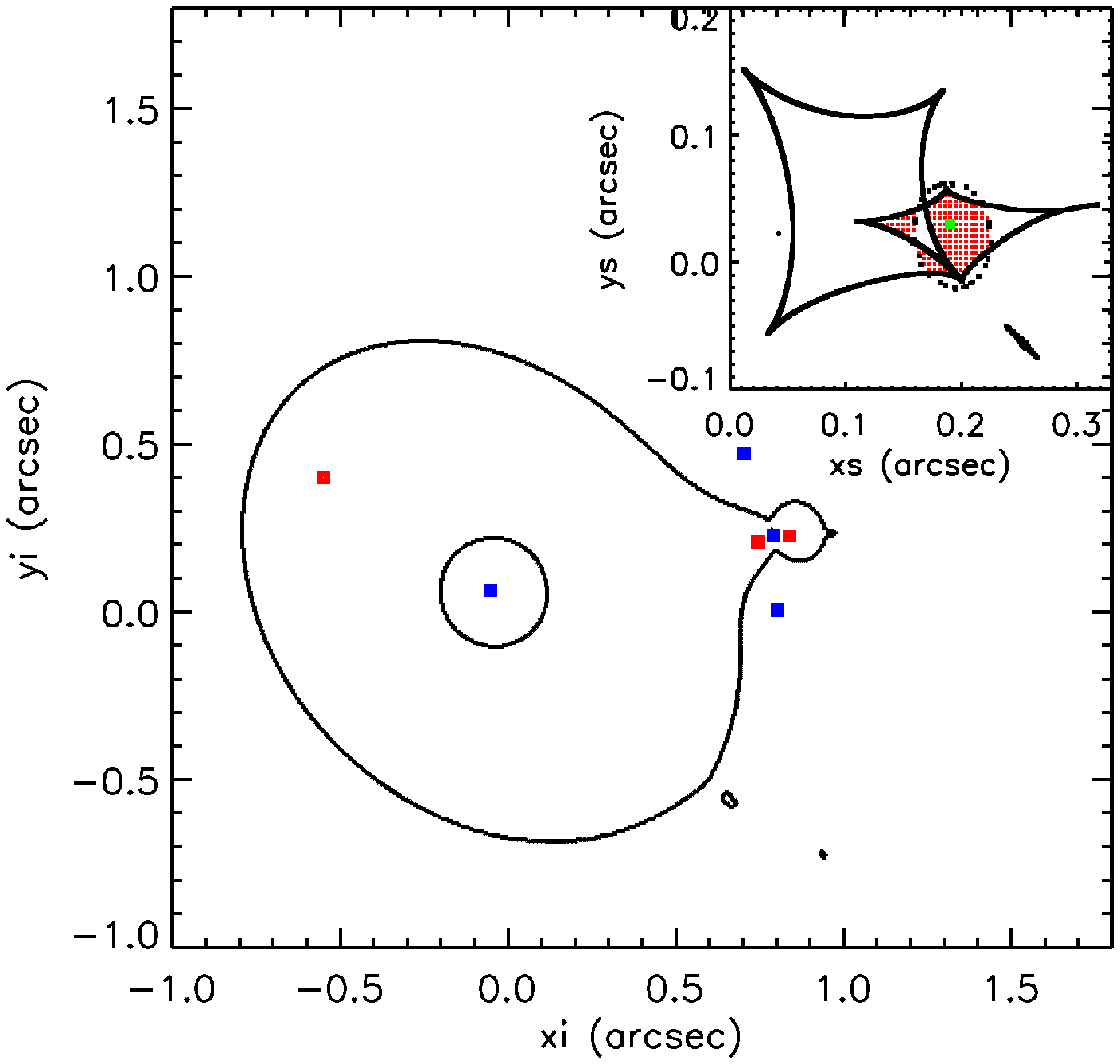}
\includegraphics[scale=0.3]{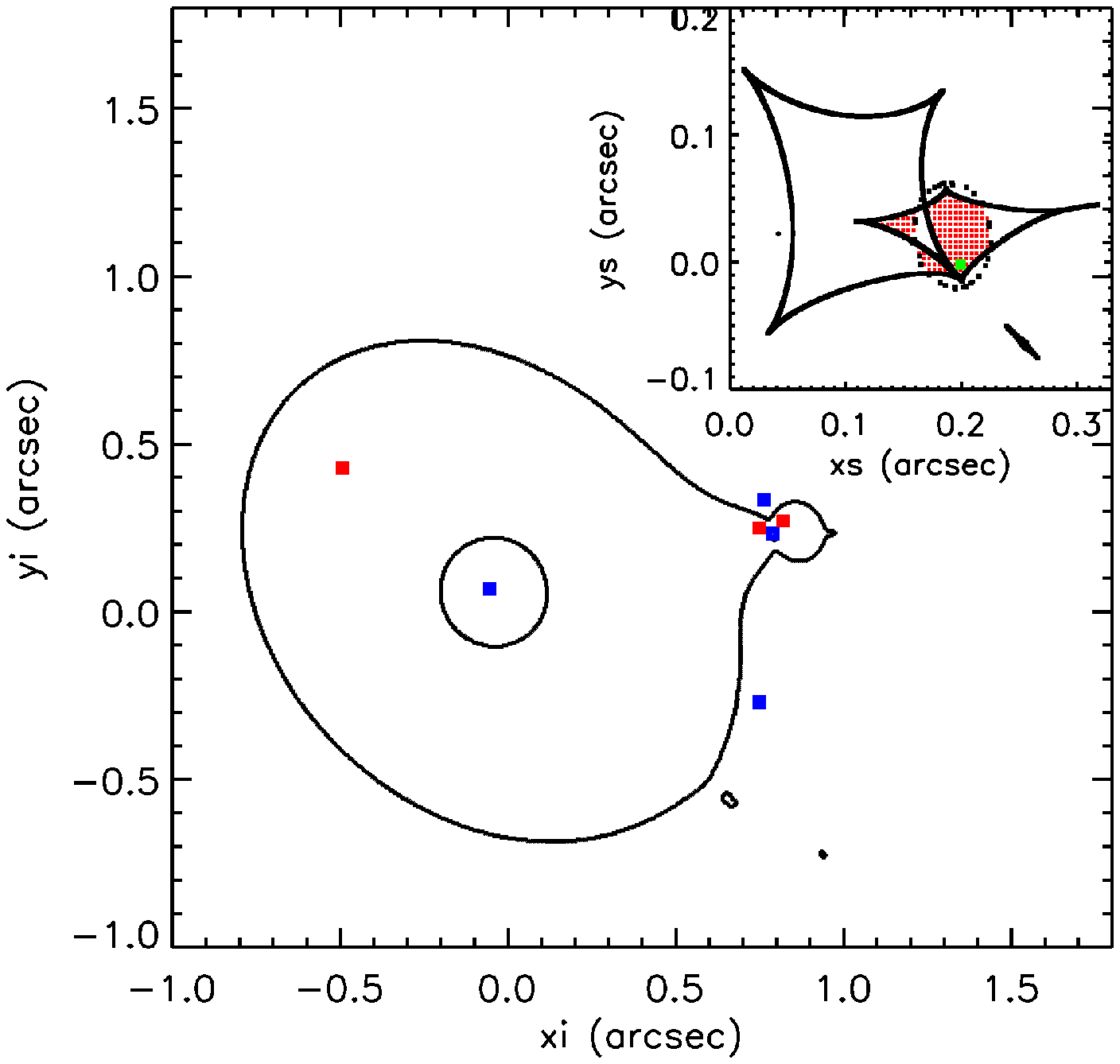}
\includegraphics[scale=0.3]{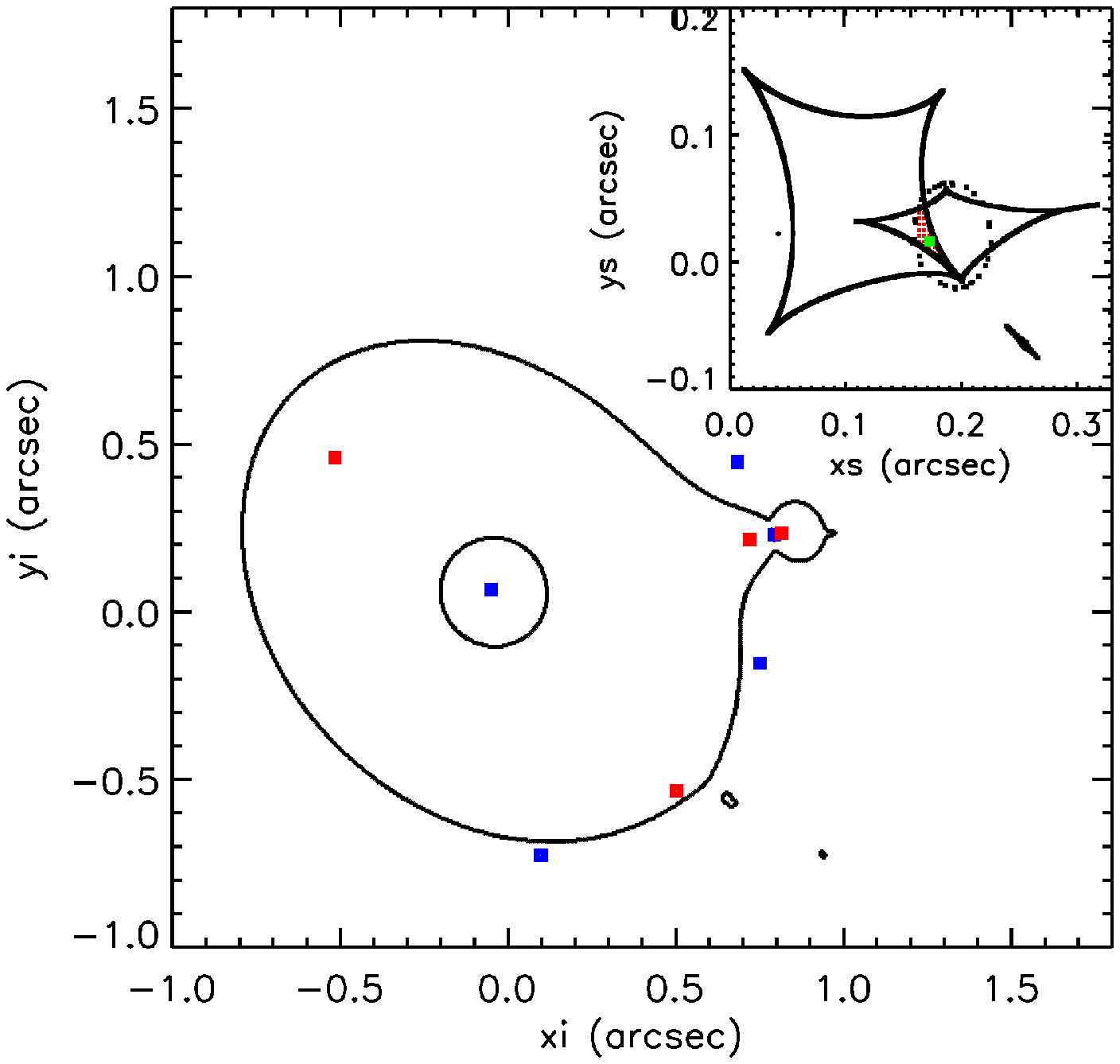}
\caption{Critical curves and caustics of an example MS-II lensing
  cone. The remarkable wiggle feature on the tangential critical curve
  is caused by a $2\times10^{10} M_{\odot}$ halo along the line of
  sight, modelled as a truncated singular isothermal sphere. Green
  dots in red regions in the subpanels are example source
  positions. Red and blue dots in the main panels are the
  corresponding image positions; red for negative parities, blue for
  positive parities. Panels 1-4 present different image configurations
  for a source located within the caustic region where five images
  would be produced; panel 5-8 are for a caustic source with seven
  images; and panel 9 is for nine images.  } 
\label{fig:LC2CaseStudy}
\end{figure*}

We have also investigated the effect from line-of-sight haloes more
massive than $10^{10} h^{-1}M_{\odot}$, which are most likely to
retain a significant fraction of baryons in their dark matter
potential wells. The chance of finding at least one of these massive
secondary lenses intercepting a strong-lensing sight line (i.e.
projected within a typical Einstein radius of 1$\arcsec$ around the
line centre, out to a redshift of 2) is about 10\%. Depending on their
density profiles, compact haloes (e.g. singular isothermal spheres)
could generate severe astrometry anomalies with a probability of a few
percent, while haloes with shallower inner profiles could not.

Fig.~\ref{fig:LC2CaseStudy} presents the peculiar image configurations
for an example sight line. Four, six and eight images (excluding the
central image) are produced when the source is located at different
positions with respect to two sets of tangential caustics. In this
particular case, the second caustic is produced by a perturbing halo
of $2\times10^{10} M_{\odot}$, modelled as a truncated singular
isothermal sphere, projected near the centre of the main lens. Such
peculiar image astrometry has already been proposed and used to
constrain density profiles of massive intergalactic objects (e.g.,
\citealt{Wyithe2001,Wilkinson2001}).

\subsection{Substructures inside line-of-sight haloes}

To investigate the effect of substructures inside haloes along the
line of sight, we have excluded all substructures from our
Millennium-II lensing cones and calculated violations due to
``smooth'' line-of-sight haloes alone. Table \ref{tab:MSIIlosRcusp}
lists violation probabilities in this case (for different halo density
profiles). The relevance of subhaloes to lensing flux-ratio anomalies
strongly depends on their assumed density profiles. Comparison of the
first and second rows of Table 2 shows that for $m > 10^{8}
h^{-1}M_{\odot}$, NFW-like substructures within line-of-sight haloes
are responsible for causing a few percent of the cusp-caustic
violations.

\subsection{Background vs. foreground}

We have separated line-of-sight haloes that are distributed in front
of and behind the main-lens plane ($z_d=0.6$). The violation
probabilities of these two groups are listed in Table
\ref{tab:MSIIlosRcusp}, calculated excluding their subhaloes.  A
higher violation probability is found caused by haloes in the
background than in the foreground, as more haloes intercept the light
rays behind the main lens plane, given a typical lensing geometry
($z_d=0.6$ and $z_s=2$). It is interesting to notice that violations
from the foreground and the background roughly add up to the total
violations due to haloes along the entire line of sight (second row of
Table \ref{tab:MSIIlosRcusp}). The ratio between violations from the
foreground and from the entire sight line is close to 2:5, which is
the ratio between the comoving radial distances for $z_d=0.6$ and
$z_s=2$.


\section{Results from Monte-Carlo haloes with a Sheth-Tormen ~~~~~~~~~~~~~~~~~~~~mass function}

The Millennium-II simulation has a limited mass resolution. To
investigate the mass dependence of the violation pattern below the
halo mass of $10^{8} h^{-1}M_{\odot}$, we have used a Monte-Carlo
method to generate intergalactic halo populations with masses $10^{6}
h^{-1}M_{\odot} \leqslant m < 10^{12} h^{-1}M_{\odot}$ (see $\S$7 for
discussion on adopting $10^{6} h^{-1}M_{\odot}$ as the lower mass
limit). These haloes are drawn from the Sheth-Tormen mass function
(\citealt{STMassFunction2002}) generated with the code provided by
\citet{ReedMassFunction2007}.

We have randomly generated 200 lensing cones out to $z_s=2$, each of
which contains a main lensing halo modelled as an isothermal ellipsoid
at redshift $z_d=0.6$. The lensing strength $b_{\rm SIE}$ is fixed to
be $0.84\arcsec$, the same as the mean $b_{\rm SIE}$ of the main
lenses in the selected sample of the Millennium-II lensing cones. The
axis ratio $q_3=0.8$, core radius $S_0=0.05\arcsec$ and an orientation
angle of 0.25$\pi$ are also taken to be the same for all main lenses.

\begin{table}
  \centering \caption{Mean surface number densities of projected haloes
    out to $z=2.0$ per decade of mass, averaged over 200 Monte-Carlo lensing
    cones.} \small\addtolength{\tabcolsep}{-1.5pt}
\label{tab:MChaloSurfaceNumber}
\begin{minipage} {\textwidth}
\begin{tabular}[b]{c|c|c|c|c}\hline
$[10^{6}, 10^7] $ & $[10^{7}, 10^8]$ & $[10^{8}, 10^9]$ & $[10^{9}, 10^{10}]$ & $ \geqslant 10^{10}$ ($h^{-1}M_{\odot}$) \\
\hline 414 & 50 & 6 & 0.7 & 0.1 (/arcsec$^2$) \\
\hline
\end{tabular}
\end{minipage}
\end{table}

In each realization of the lensing cone, line-of-sight halo positions
are randomly generated, with number densities as given by the
Sheth-Tormen mass function at the redshifts of the 60 lens planes used
for the Millennium-II lensing
cones. Table~\ref{tab:MChaloSurfaceNumber} lists the mean surface
number densities of projected haloes in different mass decades,
averaged over 200 lensing cones\footnote{These numbers roughly follow
  a power-law mass function of $dn(m)=m^{-1.9}dm$, $m$ being the mass
  of haloes.}. Haloes projected within the $50\arcsec\times
50\arcsec$-cone are saved, and those projected within the central
$5\arcsec\times 5\arcsec$-region are modelled with truncated singular
isothermal spheres and truncated NFW profiles (using both M08 and
B01-M05 concentration-mass relations). Those further out are modelled
with point masses. Cusp-caustic violations were identified in the same
way as for the Millennium-II lensing cones.

\subsection{Mass dependence of the cusp-caustic violation}

\begin{table*}
\centering \caption{Violation probabilities for Monte-Carlo lensing
cones with main lens parameter $b_{\rm SIE}=0.84\arcsec$.}
\small\addtolength{\tabcolsep}{-1.5pt} \label{tab:MCPRcusp}
\begin{minipage} {\textwidth}
\begin{tabular}[b]{l|c|c|c|c|c}\hline
$P^{90}(\Rcusp^{0.187})$ & ~$ \geqslant 10^{6} h^{-1}M_{\odot}$~ & ~$ \geqslant 10^7 h^{-1}M_{\odot}$~ & ~$\geqslant 10^8 h^{-1}M_{\odot}$~ & ~$\geqslant 10^{9} h^{-1}M_{\odot}$~ & ~$\geqslant 10^{10} h^{-1}M_{\odot}$~ \\
\hline
truncated singular isothermal sphere & 8.2\% & 7.7\% & 6.5\% & 4.8\% & 2.8\% \\
truncated NFW profile (M08) & 12\% & 9.1\% & 5.6\% & 2.5\% & $<1\%$ \\
truncated NFW profile (B01-M05) & 23\% & 18\% & 11\% & 5.5\% & 1.9\% \\
\hline
\end{tabular}
\end{minipage}
\end{table*}

The lower panels of Fig.~\ref{fig:losvsmainSISNFW} show $P(\geqslant
\Rcusp | \Delta\theta \pm2.5^{\circ})$ contour maps using Monte-Carlo
realizations of the line-of-sight halo population, with a lower mass
limit of $10^6 h^{-1}M_{\odot}$. As in the MS-II lensing case, the
frequency of cusp-caustic violations depends strongly on the assumed
halo density profiles. Table~\ref{tab:MCPRcusp} also presents the
values of $P^{90}(\Rcusp^{0.187})$ when this lower mass limit is
increased, so that the mass dependence of the cusp-caustic violations
can be seen.

Applying truncated singular isothermal spheres yields relatively
larger contributions to cusp violations from more massive
haloes\footnote{\citet{Dandan09AquI} estimated the total lensing
  cross-section $\sigma_{\rm cs} \propto b_{\rm SIE}^2 \times N_{\rm
    perturbers} (m)$ for singular isothermal lenses. In this case
  $\sigma_{\rm cs} \propto m^{\alpha}$, and $\alpha$ is a positive
  value, hence the total lensing cross-section will be biased towards
  massive haloes.} ($m \geqslant 10^{9\sim10} h^{-1}M_{\odot} $). When
truncated NFW profiles are assumed, lower mass haloes would also cause
a significant amount of violations. As can be seen from
Table~\ref{tab:MCPRcusp}, when the lower mass limit of line-of-sight
haloes decreases from $10^8 h^{-1}M_{\odot}$ to $10^6 h^{-1}M_{\odot}$
($10^7 h^{-1}M_{\odot}$), the corresponding violation probabilities,
$P^{90}(\Rcusp^{0.187})$, increase by $\sim$2\% (1\%), $\sim$6\% (4\%)
and 12\% (7\%) when assuming truncated singular isothermal spheres and
truncated NFW profiles with the M08 and the B01-M05 concentration-mass
relations, respectively.

Comparing Table~\ref{tab:MCPRcusp} with Table~\ref{tab:MSIIlosRcusp},
it can be seen that our Monte-Carlo results are similar to those
obtained using the Millennium-II lensing cones in the mass range above
$10^8 h^{-1}M_{\odot}$. This suggests that the clustering of haloes is
not a dominant effect in the production of flux-ratio anomalies for
galactic-scale lenses.

\subsection{Dependence on the Einstein radius}

\begin{table*}
\centering \caption{$P^{90}(\Rcusp^{0.187})$ for Monte-Carlo lensing
cones with main lenses of different Einstein radii $b_{\rm SIE}$:
violations due to line-of-sight perturbers more massive than $10^6
h^{-1}M_{\odot}$.} \small\addtolength{\tabcolsep}{-1.5pt}
\label{tab:MCbIRcusp}
\begin{minipage} {\textwidth}
\begin{tabular}[b]{l|c|c|c|c}\hline
$P^{90}(\Rcusp^{0.187})$ & $b_{\rm SIE}=0.62\arcsec$~ & $b_{\rm SIE}=0.84 \arcsec$ & ~$b_{\rm SIE}=1.0 \arcsec$~ & ~$b_{\rm SIE}=1.5 \arcsec$~ \\
\hline
truncated singular isothermal sphere & 6.4\% & 8.2\% & 11\% & 15\% \\
truncated NFW profile (M08) & 9.5\% & 12\% & 15\% & 20\% \\
truncated NFW profile (B01-M05) & 20\% & 23\% & 28\% & 32\% \\
\hline
\end{tabular}
\end{minipage}
\end{table*}

When the lower mass cutoff for main lensing haloes in the
Millennium-II lensing cones is reduced from $10^{12}
h^{-1}M_{\odot}$ to $10^{11} h^{-1}M_{\odot}$, the mean lensing
strength $b_{\rm SIE}$ of the equivalent isothermal ellipsoids
decreases from $0.84\arcsec$ to $0.62\arcsec$ (corresponding to
$\sigma_{\rm SIE}= 190\kms$ for $z_d=0.6$ and $z_s=2$ using
Eq.\ref{eq:V200SigmaSIE}). Table~\ref{tab:MCbIRcusp} presents the
$P^{90}(\Rcusp^{0.187})$ values that result from four different
$b_{\rm SIE}$ for the main lenses in our Monte-Carlo lensing
cones. In addition to $b_{\rm SIE }=0.62\arcsec$ and
$0.84\arcsec$, we have calculated violations for an arbitrary
$b_{\rm SIE}$ of $1\arcsec$ ($1.5\arcsec$), which is about the mean
(largest) Einstein radius of the observed systems listed in
Table~\ref{tab:obs120}.

As can be seen clearly from Table~\ref{tab:MCbIRcusp}, systems with
larger Einstein radii have higher violation probabilities. This is
expected, because close triple images normally form around the
tangential critical curve at about the Einstein radius. Comparing to
the case of a small Einstein radius, a larger value of this radius
results in a higher chance for the image triple (of a given opening
angle $\Delta\theta$) to be intercepted by line-of-sight perturbers.

\subsection{Halo concentrations and mass function}

\begin{figure}
\includegraphics[width=7cm]{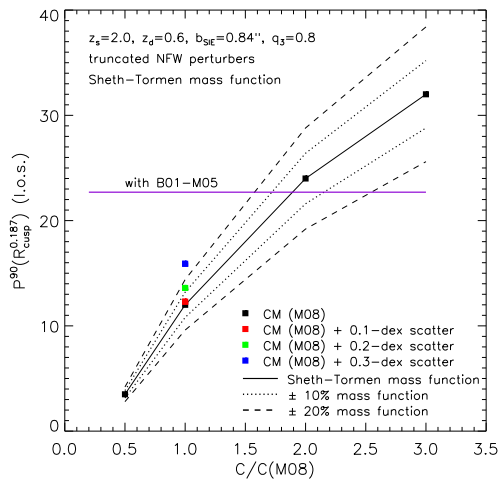}

\caption{The violation probability $P^{90}(\Rcusp^{0.187})$ changes
  with the application of different concentration-mass relations.
  $P^{90}(\Rcusp^{0.187})$ is given by the Y-axis. The X-axis
  indicates the assumed halo concentration $C$ (at any given mass)
  normalized to $C$(M08) -- the concentration predicted by the M08
  concentration-mass relation. The Sheth-Tormen mass function is used
  to generate line-of-sight halo populations. Values of
  $P^{90}(\Rcusp^{0.187})$ at $C/C$(M08)=0.5, 1.0, 2.0 and 3.0 are
  plotted as the four black squares, which are connected by the black
  solid line. Assuming that violations grow linearly with the number
  of perturbers, the dash and dotted lines plotting
  $(100\pm20)\%\times$ and $(100\pm10)\%\times$ the
  $P^{90}(\Rcusp^{0.187})$ values as shown in the black solid line,
  resemble violation probabilities under $(100\pm20)\%\times$ and
  $(100\pm10)\%\times$ the Sheth-Tormen mass function,
  respectively. Red, green and blue squares present results (under the
  Sheth-Tormen mass function) when allowing for scatter of
  concentrations (for haloes of a given mass) in form of Gaussian
  distributions with the standard deviation being 0.1, 0.2 and 0.3 dex
  around mean concentration values predicted by the M08
  concentration-mass relation. $P^{90}(\Rcusp^{0.187})$ derived from
  using the B01-M05 concentration-mass relation is also given,
  indicated by the purple horizontal line.}
 \label{fig:MoPlot}
\end{figure}

As we have shown above, the cusp-violation probability depends
strongly on our assumptions about halo concentration. The
concentration-mass relation derived by Bullock et al. (2001) has a
simple functional form (including redshift evolution) and has been
widely used in the literature. \citet{ColinNFWC2004} investigated
concentration parameters for haloes of $ 10^{6} h^{-1}M_{\odot}
\leqslant m \leqslant 10^{9} h^{-1}M_{\odot}$ and found this
relationship to be a good fit. However, these early simulations of
dark matter halo formation had relatively low numerical resolution
and this can introduce systematic errors.

More recently, a number of authors including \citet{Neto07CM},
\citet{Gao08CM}, \citet{Zhao09CM} and M08 (whose results are used
above), derived concentration-mass relations from high-resolution
cosmological $N$-body simulations. These studies are restricted to
haloes with $m \geqslant 10^{10} h^{-1}M_{\odot}$ but they exhibit
systematic differences from the concentrations obtained by
B01-M05. For this reason we show lensing results using the M08
relation, extrapolating to lower masses when required, but exploring
how the results change when this relation is varied by factors of a
few.

The B01-M05 concentration-mass relation overestimates the
concentration of small mass haloes inferred from the extrapolated M08
relation by factors of $3\sim 4$ at $z=0$. Therefore, we expect the
violation probability to be larger for the B01-M05 relation than for
the M08 relation.  The scatter in halo concentration also affects the
final cusp-violation probability.

Fig.~\ref{fig:MoPlot} presents the values of $P^{90}(\Rcusp^{0.187})$
induced by line-of-sight haloes assuming the Sheth-Tormen mass
function. To allow for possible uncertainties in halo concentration,
we also show results for the case when the concentrations inferred
from the M08 concentration-mass relation are multiplied by factors of
0.5, 1.0. 2.0 and 3.0. Varying amounts of scatter in concentration
(for haloes of a given mass) are modelled assuming a Gaussian
distribution with mean value equal to the M08 concentrations. As may
be seen, the violation probabilities depend strongly on halo
concentrations. Higher concentrations result in higher cusp-caustic
violation probabilities. A larger scatter in concentration will also
increase the violation probability.

The halo mass function (the number density of haloes per unit volume
per decade in mass) influences the cusp violation probability.
\citet{Metcalf2005b} found that flux-ratio anomalies caused by
line-of-sight perturbers not only depend strongly on the radial
density profile of the haloes (their concentration), but also on the
primordial matter power spectrum on small scales.
\citet{Miranda2007} suggested that flux-ratio anomalies could be
used statistically as a test of the behaviour of the matter power
spectrum on small scales.  We do not explore these effects here but in
Fig.~\ref{fig:MoPlot} we show the result of using
$(100\pm10/20)\%\times$ the Sheth-Tormen mass function, and assuming
that violations grow linearly with the number of perturbers.

\section{DISCUSSION AND CONCLUSIONS}

We have examined the effects of intergalactic cold dark matter haloes
on flux-ratio anomalies in multiply-lensed quasar images by
ray-tracing along strong-lensing sight lines that are either taken
from the Millennium II simulation (for haloes and subhaloes with
$m>10^8 h^{-1}M_{\odot}$), or generated using the Monte-Carlo method
assuming a Sheth-Tormen mass function (for haloes with $m\geqslant10^6
h^{-1}M_{\odot}$).

We use $P^{90}(\Rcusp^{0.187})$, the probability for the cusp-caustic
relation, $\Rcusp$, to be larger than or equal to 0.187 -- the
smallest value of $\Rcusp$ measured for cusp-caustic systems to date
(for the quasar B1422) -- over all realizations with $\Delta\theta
\leqslant 90^{\circ}$, as a statistical measure of the cusp-caustic
violation probability. We have found that the mean violation
probability from intervening haloes depends strongly on their density
profiles. 

\citet{Chen2003} assumed singular isothermal spheres for line-of-sight
haloes and find that they only contribute to $\leqslant 10 \%$ of the
total perturbation. Assuming the same halo density profile, we find
that the cusp-caustic violation probability caused by line-of-sight
haloes with $m\geqslant 10^{6} h^{-1}M_{\odot}$ is comparable to that
caused by intrinsic substructures within the main lensing halo
($P^{90}(\Rcusp^{0.187})\approx8\%$ vs. $10\%$,
\citealt{Dandan2010AqII}), which is in good agreement with
\citet{Miranda2007}. The different results between \citet{Chen2003}
and ours can be attributed to the drawbacks of their cross-section
method, which underestimates effects from more sophisticated
perturbation scenarios (see \citealt{Metcalf2005a}).

When we assume truncated NFW profiles for the line-of-sight haloes
($m\geqslant 10^{6} h^{-1}M_{\odot}$), the violation probability,
$P^{90}(\Rcusp^{0.187})$, increases to $23\%$ if we adopt the B01-M05
concentration-mass relation and to $12\%$ if we adopt our preferred
relation, that by M08. These values are larger than that due to the
intrinsic subhalo populations alone.

A typical NFW profile has an Einstein radius $3\sim4$ orders of
magnitude smaller than a singular isothermal sphere with a same
mass. However, NFW perturbers in the mass range from $10^{6}
h^{-1}M_{\odot}$ to $10^{9\sim 10} h^{-1}M_{\odot}$ cause more cusp
violations than their singular isothermal counterparts. This may be
due to the fact that in this mass range, perturbation in magnification
(ratios) is mainly from fluctuations in the local density field that
do not change the image positions. When comparing an NFW with a
singular isothermal sphere of the same mass, we notice that the
surface density distribution of the former exceeds that of the latter
from a radius of $\sim0.001r_{200}$ outwards, which means the NFW
profile is more effective in introducing fluctuation to the
convergence and thus causing flux-ratio anomalies.

On the other hand, the deflection angle of a perturber of $m\sim
10^{6-9} h^{-1}M_{\odot}$ is always small ($\lesssim 0.001\arcsec$ for
a singular isothermal sphere locating at $z=0.6$), until the perturber
is massive ($m \gtrsim 10^{10} h^{-1}M_{\odot}$) and compact enough (a
singular isothermal sphere) to have a deflection angle ($\gtrsim
0.01\arcsec$) that can shift a nearby image to a new position with a
different magnification from the primary lens (see
\citealt{Metcalf2005b}). This can explain the larger violation
probabilities (as shown in Table 4), caused by singular isothermal
perturbers of $m\geqslant10^{9\sim10} h^{-1}M_{\odot}$ than by their
NFW counterparts which are less effective in causing flux-anomalies
due to shifting image positions. \\

Another issue concerns the finite-source
effect. \citet{Metcalf2010FluxAnomaly} pointed out that biased results
about substructures could be drawn due to the point source
approximation, which is used in this work.

The radio-emission regions of observed quasars are estimated to be
$\sim$10 parsecs in extent (\citealt{Andreani1999,Wyithe2002}),
corresponding to $\sim0.001\arcsec$ for a source at $z_s=2.0$.  When
the perturbing mass drops down below $10^6 h^{-1}M_{\odot}$, the
corresponding effective perturbing area decreases to
$\lesssim0.001\arcsec$ in radius for the perturber at $z_d=0.6$,
becoming smaller than an image with $\mu\sim10-20$ (around the
tangential curve) of the radio emission region of a background
quasar. As a result, the induced magnification fluctuation would be
smeared out (within the image area), and thus no significant image
flux anomaly would be observed at radio wavelengths (but could still
be seen in the optical/near-infrared, which comes from much smaller
physical regions. See \citealt{Moustakas2003} for spectroscopic
gravitational lensing).
This is why we do not consider the violation probability produced by
perturbing haloes below $10^6 h^{-1}M_{\odot}$. As can be seen from
Table 4, even if we neglect contributions from perturbers below $10^7
h^{-1}M_{\odot}$, we still find $\sim10\%$ cusp-violation probability
from line-of-sight NFW-like perturbers adopting the M08
concentration-mass relation. \\

Several other points are worth noting. Firstly, the violation probability
depends, of course, on the concentration of the halo, and both large
halo concentrations and a large scatter in concentration will result
in higher violation probabilities. Thus, it may be possible to use the
statistics of flux-ratio distributions (measured in the radio) from
large samples of lensed quasars to constrain the density profiles of
low-mass dark matter haloes.

Secondly, in \citet{Dandan09AquI} we found that the violation
probability is higher for systems with larger Einstein radii because
the mass fraction in dark substructures increases with radius.  Here,
the probability of a system violating the cusp-caustic relation is
also seen to increase with the Einstein radius but for a different
reason: close triple images (with a given opening angle
$\Delta\theta$) that form at larger radii are more likely to be
intercepted by line-of-sight perturbers. Adopting $b_{\rm SIE
}=1.0\arcsec$, the mean Einstein radius for the observed sample, we
find that the violation probability $P^{90}(\Rcusp^{0.187})$ increases
to 15\% if the M08 concentration-mass relation is adopted. We also
note that if we use the B01-M05 concentration-mass relation, the
violation probabilities for the representative cases in
\citet{Metcalf2005a} can be reproduced.

Third, the $\Rcusp$-$\Delta\theta$ distribution varies with the
ellipticity of the main lens in such a way that lenses with higher
ellipticity (smaller axis ratio $q_3$) have larger $\Rcusp$, and thus
potentially larger $P^{90}(\Rcusp^{0.187})$ (\citealt{KGP2003apj,
  Metcalf2010FluxAnomaly}). In this work, we adopt $q_3=0.8$ for the
simulated main lensing haloes. For four of the five lenses shown in
Table 1 (including all three with image opening angle $\Delta\theta
\leqslant 90^{\circ}$), the axis ratios (0.75--0.9 as inferred from
the singular isothermal ellipsoid models, see \S3.1) are close to the
value we adopt (0.8) in our simulations. In the absence of perturbing
structures, changing the axis ratio to $q_3=0.7$ causes a negligible
increase in $P^{90}(\Rcusp^{0.187})$.

Fourth, the probability that a massive halo ($m\geqslant10^{10}
h^{-1}M_{\odot}$) intercepts a galaxy-scale strong-lensing sight line
with an impact parameter of $\leqslant 1\arcsec$ from the main lens
centre is about 10\%.  These halos can perturb image fluxes, surface
brightness (e.g., \citealt{VKBTG2009}) and even image astrometry
(e.g., \citealt{Wyithe2001}). Halos with compact density profiles
(e.g., singular isothermal spheres) could generate extra image pairs
locally with an image separation of $0.01\arcsec\sim0.1\arcsec$,
resulting in more than four bright images of a background quasar. With
upcoming lensing surveys, observations of peculiar image
configurations could put better constraints on the density profiles of
these massive haloes (\citealt{XivryMarshall09}).

Finally, for masses above $10^{8} h^{-1}M_{\odot}$, we find that halo
clustering has only a minor effect on the flux-ratio anomalies for
galaxy-scale lensing systems. For a typical lensing geometry (with
$z_d=0.6$, $z_s=2$), the overall perturbation produced by background
haloes (behind the main lens) is larger than that caused by foreground
haloes.  \\


To summarise, in this work we have calculated the cusp-caustic
violation probability, as measured by $P^{90}(\Rcusp^{0.187})$,
produced by line-of-sight dark matter haloes. The value of
$P^{90}(\Rcusp^{0.187})$ strongly depends on halo density profiles,
specifically on concentration, in the case of NFW perturbers. When the
concentration-mass relation proposed by \citet{Maccio08CM} is used,
the value of $P^{90}(\Rcusp^{0.187})$ produced by all line-of-sight
perturbers is found to be $\sim10-15\%$ (corresponding to a mean
Einstein radius of $0.8-1.0\arcsec$). In previous work
(\citealt{Dandan09AquI,Dandan2010AqII}) using the Aquarius simulations
(\citealt{volker08Aq}), we found that the contribution to
$P^{90}(\Rcusp^{0.187})$ from substructures inside the main lensing
haloes also amounts to $P^{90}(\Rcusp^{0.187})\sim 10-15\%$
(corresponding to a mean Einstein radius in the same range as
above). Summing up both contributions, the total violation probability
could reach $\sim20-30\%$.

There are five observed cusp-geometry lensing systems whose triple
images have opening angles $\Delta\theta \leqslant 90^{\circ}$. Of
these five, the three radio lensing cases (B0712+472, B1422+231 and
B2045+265 as listed in Table 1) show firm evidence for cusp-caustic
violations caused by galactic-scale structures. Applying the same
statistical argument as \citet{Dandan09AquI}, we conclude that the
chance of observing such a violation rate (3/5) is $\sim 5-13\%$ for a
total $P^{90}(\Rcusp^{0.187})$ of $\sim 20-30\%$. This can be compared
with the chance of $\la1\%$ that we found when considering only
intrinsic substructures
(\citealt{Dandan09AquI,Dandan2010AqII}). However, we caution that such
a simple statistic argument does not take into account the full
two-dimensional probability distribution in the
$\Rcusp$-$\Delta\theta$ plane.

The existing observational sample is clearly too small for us to reach
a definitive conclusion regarding the appropriateness of the
$\Lambda$CDM model. Our main result, however, is that, depending on
the density profiles of CDM haloes (and subhaloes), the line-of-sight
projection effect on the flux-ratio anomalies of quasar images can be
comparable to or even larger than that from intrinsic subhalos. The
resulting cusp-violation probability from the combined effect
alleviates the discrepancy between the CDM model and current data.
New multiply-lensed four-image systems discovered in upcoming lensing
surveys will make it possible to use the statistics of flux-ratio
anomalies to constrain the properties of dark matter halo as well as
the cosmogonic model.

We end by noting that a warm dark matter cosmogony has a different
power spectrum of density perturbations, as well as different density
profiles for small halos compared to the standard CDM cosmogony. This
will result in different (presumably lower) cusp-violation
probabilities (e.g., \citealt{Miranda2007}). This possibility is worth
exploring further in future.

\section*{Acknowledgements} We thank Jie Wang, Houjun Mo, Leon
Koopmans, Anna Nierenberg, Peter Schneider, Stefan Hilbert and
Dominique Sluse for helpful and insightful discussions. We also thank
the referee for helpful suggestions. Thanks also go to Dr. Lydia Heck
for her skillful management of and persistent dedication to the
computer clusters at the Institute for Computational Cosmology in
Durham, where the lensing simulations were carried out. SM
acknowledges financial support from the Chinese Academy of Sciences,
LG acknowledges support from an STFC advanced fellowship and a
one-hundred-talents program of the Chinese Academy of Sciences (CAS)
and the National Basic Research Program of China (973 program under
grant No. 2009CB24901). CSF acknowledges a Royal Society Wolfson
Research Merit award and an ERC advance investigator grant. REA
acknowledges support from the Advanced Grant 246797 "GALFORMOD" from
the European Research Council. This work was supported in part by an
STFC rolling grant to the ICC.

\bibliographystyle{mn2e}
\bibliography{ms_xudd}
\label{lastpage}

\end{document}